\documentclass[sn-basic]{sn-jnl}

\usepackage{bm}
\usepackage{color}
\definecolor{red}{rgb}{1.0, 0.0, 0.0}

\jyear{2022}%

\begin{document}

\title[Reconnection in the lab]{Laboratory Study of Collisionless Magnetic Reconnection}


\author*[1,2]{\fnm{H.} \sur{Ji}}\email{hji@pppl.gov}

\author[2]{\fnm{J.} \sur{Yoo}}

\author[2]{\fnm{W.} \sur{Fox}}

\author[2]{\fnm{M.} \sur{Yamada}}

\author[3]{\fnm{M.} \sur{Argall}}

\author[4]{\fnm{J.} \sur{Egedal}}

\author[5]{\fnm{Y.-H.} \sur{Liu}}

\author[6]{\fnm{R.} \sur{Wilder}}

\author[7]{\fnm{S.} \sur{Eriksson}}

\author[8]{\fnm{W.} \sur{Daughton}}

\author[1]{\fnm{K.} \sur{Bergstedt}}

\author[2]{\fnm{S.} \sur{Bose}}

\author[9]{\fnm{J.} \sur{Burch}}

\author[3]{\fnm{R.} \sur{Torbert}}

\author[10,11,2]{\fnm{J.} \sur{Ng}}

\author[11]{\fnm{L.-J.} \sur{Chen}}

\affil*[1]{\orgdiv{Department of Astrophysical Sciences}, \orgname{Princeton University}, \orgaddress{\street{4 Ivy Lane}, \city{Princeton}, \postcode{08544}, \state{New Jersey}, \country{U.S.A}}}

\affil[2]{\orgname{Princeton Plasma Physics Laboratory}, \orgaddress{\street{P.O. Box 451}, \city{Princeton}, \postcode{08543}, \state{New Jersey}, \country{U.S.A}}}

\affil[3]{\orgdiv{Institute for the Study of Earth, Oceans, and Space}, \orgname{University of New Hampshire}, \orgaddress{\street{8 College Road}, \city{Durham}, \postcode{03824}, \state{New Hampshire}, \country{U.S.A}}}

\affil[4]{\orgdiv{Department of Physics}, \orgname{University of Wisconsin - Madison}, \orgaddress{\street{1150 University Avenue}, \city{Madison}, \postcode{53706}, \state{Wisconsin}, \country{U.S.A}}}

\affil[5]{\orgdiv{Department of Physics and Astronomy}, \orgname{Dartmouth College}, \orgaddress{\street{17 Fayerweather Hill Road}, \city{Hanover}, \postcode{03755}, \state{New Hampshire}, \country{U.S.A}}}

\affil[6]{\orgdiv{Department of Physics}, \orgname{University of Texas at Arlington}, \orgaddress{\street{701 S. Nedderman Drive}, \city{Arlington}, \postcode{76019}, \state{Texas}, \country{U.S.A}}}

\affil[7]{\orgdiv{Laboratory for Atmospheric and Space Physics}, \orgname{University of Colorado at Boulder}, \orgaddress{\street{1234 Innovation Drive}, \city{Boulder}, \postcode{80303}, \state{Colorado}, \country{U.S.A}}}

\affil[8]{\orgname{Los Alamos National Laboratory}, \orgaddress{\street{P.O. Box 1663}, \city{Los Alamos}, \postcode{87545}, \state{New Mexico}, \country{U.S.A}}}

\affil[9]{\orgname{Southwest Research Institute}, \orgaddress{\street{6220 Culebra Road}, \city{San Antonio}, \postcode{78238}, \state{Texas}, \country{U.S.A}}}

\affil[10]{\orgdiv{Department of Astronomy}, \orgname{University of Maryland}, \orgaddress{\street{4296 Stadium Drive}, \city{College Park}, \postcode{20742}, \state{Maryland}, \country{U.S.A}}}

\affil[11]{\orgname{Goddard Space Flight Center}, \orgaddress{\street{Mail Code 130}, \city{Greenbelt}, \postcode{20771}, \state{Maryland}, \country{U.S.A}}}

\abstract{A concise review is given on the past two decades’ results from laboratory experiments on collisionless magnetic reconnection in direct relation with space measurements, especially by Magnetospheric Multiscale (MMS) mission. Highlights include spatial structures of electromagnetic fields in ion and electron diffusion regions as a function of upstream symmetry and guide field strength; energy conversion and partition from magnetic field to ions and electrons including particle acceleration; electrostatic and electromagnetic kinetic plasma waves with various wavelengths; and plasmoid-mediated multiscale reconnection. Combined with the progress in theoretical, numerical, and observational studies, the physics foundation of fast reconnection in colisionless plasmas has been largely established, at least within the parameter ranges and spatial scales that were studied. Immediate and long-term future opportunities based on multiscale experiments and space missions supported by exascale computation are discussed, including dissipation by kinetic plasma waves, particle heating and acceleration, and multiscale physics across fluid and kinetic scales.}


\maketitle

\renewcommand*\contentsname{Reconnection in the lab}
\tableofcontents

\section{Introduction}\label{sec1}

The history of laboratory study of magnetic reconnection goes back to 1960s~\citep[e.g.][]{bratenahl70} not long since development of the early models~\citep{sweet58,parker57,dungey61,petschek64}. As briefly reviewed by \cite{yamada10}, these early experiments were motivated by solar flares, and carried out in collision-dominated MHD regime at low Lundquist numbers ($S<10$). The subsequent landmark experiments performed by~\cite{stenzelgekelman1} were also in a largely collisional ($S<10$) but electron-only regime where ions are unmagnetized even with a strong guide field. While these experiments provided insights of rich physics of magnetic reconnection in the collisional regimes, they are not directly relevant to collisionless reconnection in space, which is the focus of this book, and thus they are not included in this short review paper.

The modern reconnection experiments began with merging magnetized plasmas~\citep{yamada90,ono93,brown99} using technologies developed during nuclear fusion research. These were followed by driven reconnection experiments in an axisymmetric geometry: Magnetic Reconnection Experiment or MRX~\citep{yamada97b}, Versatile Toroidal Facility or VTF~\citep{egedal00}, and Terrestrial Reconnection Experiment or TREX~\citep{olson16}; and in a linear geometry: Rotating Wall Experiment (RWX)~\citep{bergerson06}, Reconnection Scaling Experiment (RSX)~\citep{furno07}, and more recent Phase Space Mapping experiment (PHASMA)~\citep{shi22}. Many of these experiments were able to reach higher Lundquist number, up to $S\sim10^3$, and with magnetized ions. As a result, plasma conditions local to the reconnecting current sheets in these experiments are nearly collisionless, motivating quantitative comparisons with \textit{in-situ} measurements by spacecraft in near-Earth space as well as predictions by Particle-In-Cell (PIC) kinetic simulations. The topics on magnetic reconnection for such comparative research include kinetic structures of diffusion regions, energy conversion from magnetic field to plasma, various plasma wave activity, as well as multiscale reconnection via plasmoid instability of reconnecting current sheets. This paper concisely reviews results from these comparative research activities and highlights several recent achievements, especially in relation with Magnetospheric Multiscale (MMS) mission. Summary of magnetic reconnection research in a broader scope can be found in review papers by~\cite{zweibel09} and \cite{yamada10}, as well as in more recent reviews~\citep{yamada22,ji22}. The latter review paper especially focuses on the future development of magnetic reconnection research by emphasizing its multiscale nature.

The rest of this review is organized in the following sections: kinetic structures of reconnecting diffusion regions in Sec.2 including both ion and electron diffusion regions (IDR and EDR), reconnection energetics in Sec.3, plasma waves in Sec.4, plasmoids during reconnection in Sec.5, followed by the future outlook in Sec.6.

\section{Kinetic structures of diffusion regions}\label{sec2}

Detailed studies of magnetic reconnection based on \textit{in-situ} measurements in the laboratory and in space began with detecting kinetic structures of diffusion regions near the X-line, as the research focus was the origin of fast reconnection in collisionless plasmas. The origin of kinetic structures to support reconnection electric field in collisionless plasmas can be understood via the generalized Ohm's law,
\begin{equation}
\bm{E} + \bm{V} \times \bm{B} = \eta_s \bm{j} + \frac{\bm{j} \times \bm{B}}{en} - \frac{\bm{\nabla}p_e}{en}- \frac{\bm{\nabla}\cdot \bm{\Pi}_e}{en}- \frac{m_e}{e} \frac{d \bm{V}_e}{dt},
\label{GeneralizedOhm}
\end{equation}
where $\bm{E}$, $\bm{V}$, $\bm{B}$, and $\bm{j}$ are electric field, velocity, magnetic field, and current density, respectively; and $\eta_s$ is the Spitzer resistivity. $n$ and $\bm{V}_e$ are the electron density and fluid velocity, respectively. The full electron pressure tensor is expressed as a sum of diagonal isotropic pressure tensor and stress tensor which includes off-diagonal pressure tensor: $\bm{P}_e \equiv p_e \bm{I} + \bm{\Pi}_e$ where $\bm{I}$ is unit tensor, and $m_e$ and $e$ are electron mass and charge, respectively. The RHS of Eq.(\ref{GeneralizedOhm}) represents non-ideal-MHD electric field in diffusion regions where $\bm{V} \times \bm{B}$ diminishes while $\bm{E}$ remains large for fast reconnection. Each of these non-ideal-MHD terms is associated with a spatial structure in steady state on the corresponding scale in electromagnetic field or electron quantities.

In collisional MHD plasmas, the only non-ideal electric field is due to collisional resistivity, $\eta_S \bm{j}$, while ions and electrons are closely coupled to behave as a single fluid, moving at the MHD fluid velocity, $\bm{V}$. In contrast, collisional resistivity is negligible in collisionless plasmas where non-ideal-MHD electric field must come from other terms on the RHS of Eq.(\ref{GeneralizedOhm}). In such plasmas, ions and electrons decouple from each other as they approach the current sheet. Ions get demagnetized in a larger ion diffusion region (IDR) while electrons get demagnetized closer to the X-line in a smaller electron diffusion region (EDR). In general, the second and third terms on the RHS of Eq.(\ref{GeneralizedOhm}), $\bm{j} \times \bm{B}/{en} - \bm{\nabla}p_e/{en}$, provide non-ideal-MHD electric field in IDR depending on the guide field strength, while the last two terms are responsible for non-ideal electric field in EDR. 
Below we review the laboratory studies of kinetic structures in both IDR and EDR, in comparisons with space measurements and numerical simulations, with or without a guide field, as well as with and without symmetries between the two upstream reconnection regions.

\begin{figure}[ht]
    \centering
    \includegraphics[width=10cm]{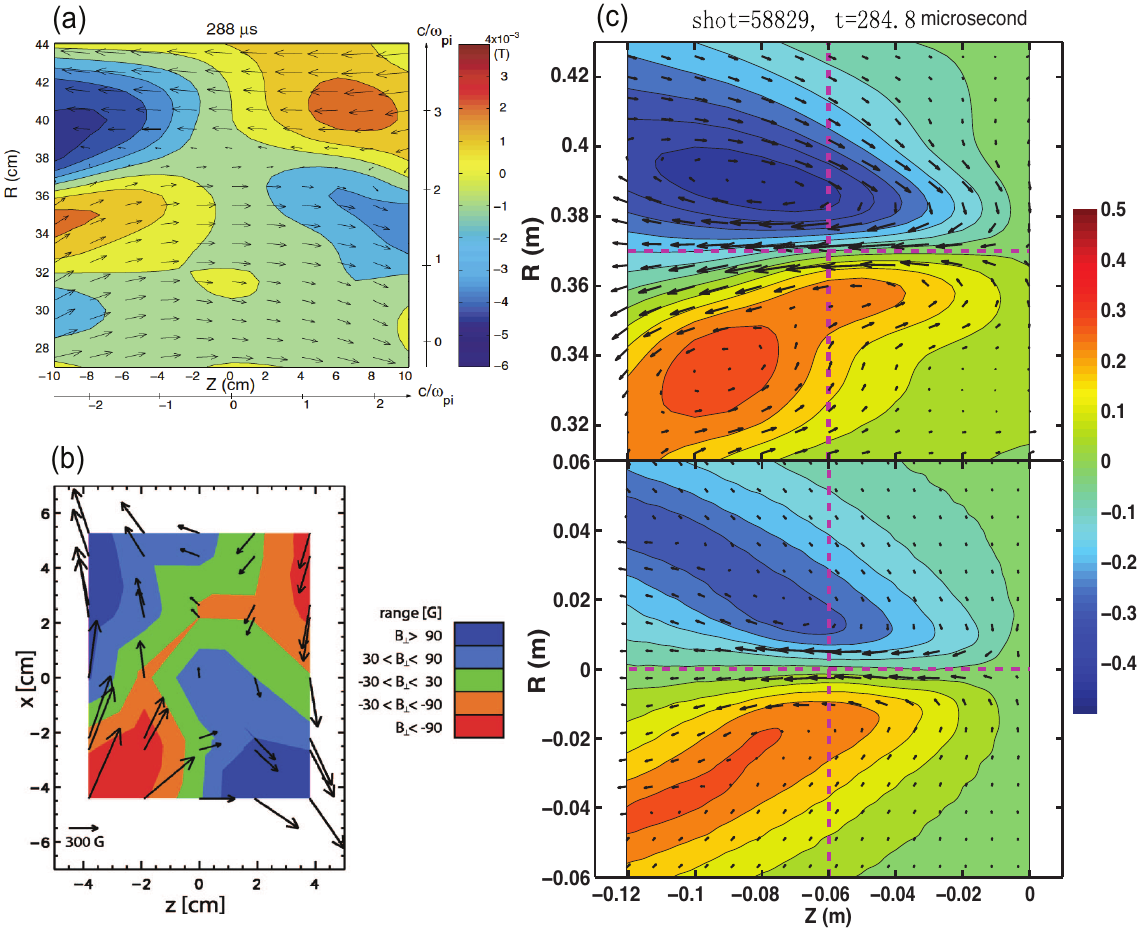}
    \caption{Measured instantaneous quadrupolar structure of out-of-the-plane magnetic field component during anti-parallel collisionless reconnection. (a) data from Magnetic Reconnection eXpriment or MRX~\citep{ren05} where $R$ is the direction across current sheet and $Z$ is along the reconnecting magnetic field; (b) data from Swarthmore Spheromak eXperiment or SSX~\citep{brown06} where $X$ is the direction across current sheet and $Z$ is along the reconnecting magnetic field; (c) comparison between MRX data (top panel) and 2D PIC simulation using corresponding parameters (bottom panel) in one half of the reconnection plane showing excellent agreements on ion scales~\citep{ji08}. Arrows indicate electron flow velocity.}
    \label{IDR1a}
\end{figure}

\subsection{IDR structures without a guide field}

When the guide field is negligible, reconnection electric field, $E_y$, is perpendicular to magnetic field which is mostly within the reconnection plane of $(z,x)$. A natural candidate to generate the required non-ideal-MHD electric field perpendicular to local magnetic field is the second term on the RHS of Eq.(\ref{GeneralizedOhm}),  $\bm{j} \times \bm{B}/{en}$, which is often called Hall term originated from the differences in the in-plane ion and electron motions as expected in IDR. Since such motions preserve symmetry between both upstreams and also both downstreams (unless distant asymmetries are imposed; see below), a quadrupolar structure in out-of-the-plane (Hall) magnetic field component, $B_y$, on the ion skin depth has been predicted theoretically~\citep{sonnerup79,terasawa83} and numerically~\citep[][and references therein]{birn01}. 

In addition to the inductive reconnection electric field in the out-of-the-plane direction, $E_y$, there may exist an in-plane electric field, $\bm{E}_\text{in-plane}$. At the outer scales (regions outside of IDR) where ideal MHD applies, the RHS of Eq.(\ref{GeneralizedOhm}) vanishes, resulting in $\bm{E}_\text{in-plane} = - (\bm{V} \times \bm {B})_\text{in-plane}$. Without a guide field, $E_\text{in-plane} = - V_yB$ which vanishes unless there exists a significant out-of-the-plane flow, $V_y$.

However, a significant $E_\text{in-plane}$, called the Hall electric field, arises even without an ion flow $V_y$ in the IDR. This is because in IDR only ions are dissipative and electrons are ideal. Therefore, $\bm{E} \approx -\bm{V_e} \times \bm {B}$ and $E_\text{in-plane} \approx - V_{ey}B \approx j_yB/en$. It also follows that $\bm{E}\cdot\bm{B} \approx 0$, and without a guide field, $\bm{E}_\text{in-plane}\cdot \bm{B} \approx 0$. In other words, $\bm{E}_\text{in-plane}$ is perpendicular to local magnetic field everywhere, which by symmetry must have a quadrupolar structure around X-line, consistent with numerical predictions~\citep[e.g.][]{shay98a}. By the virtue of Faraday's Law in quasi-steady state ($\partial B_y/\partial t \approx 0$), $\bm{E}_\text{in-plane}$ is curl-free and can be well represented by an electrostatic potential, $\bm{E}_\text{in-plane} \approx -\bm{\nabla} \phi$. Therefore $\phi$ must have a saddle-type quadrupolar structure determined by the significant out-of-the-plane $j_y$ in IDR. The presence of both $\phi$ and $B_y$ in the IDR enables fast reconnection by diverting a significant amount of incoming magnetic energy directly towards downstream in the outflow direction via Poynting vector $E_xB_y/\mu_0$, without having to pass through the X-line. Note here that $E_x$ is part of $\bm{E}_\text{in-plane}$ and peaks along the separatrix with a width on electron scales while extends to the ion scales~\citep{chen08b}. Over time, the depleted total pressure at the X-line pulls in more upstream magnetic pressure leading to the open-outflow geometry necessary for fast reconnection~\citep{liu22}. The prediction of both Hall magnetic and electric fields motivated an intensive search of such field structures as first evidence of fast collisionless reconnection.

\begin{figure}[t]
    \centering
    \includegraphics[width=5cm]{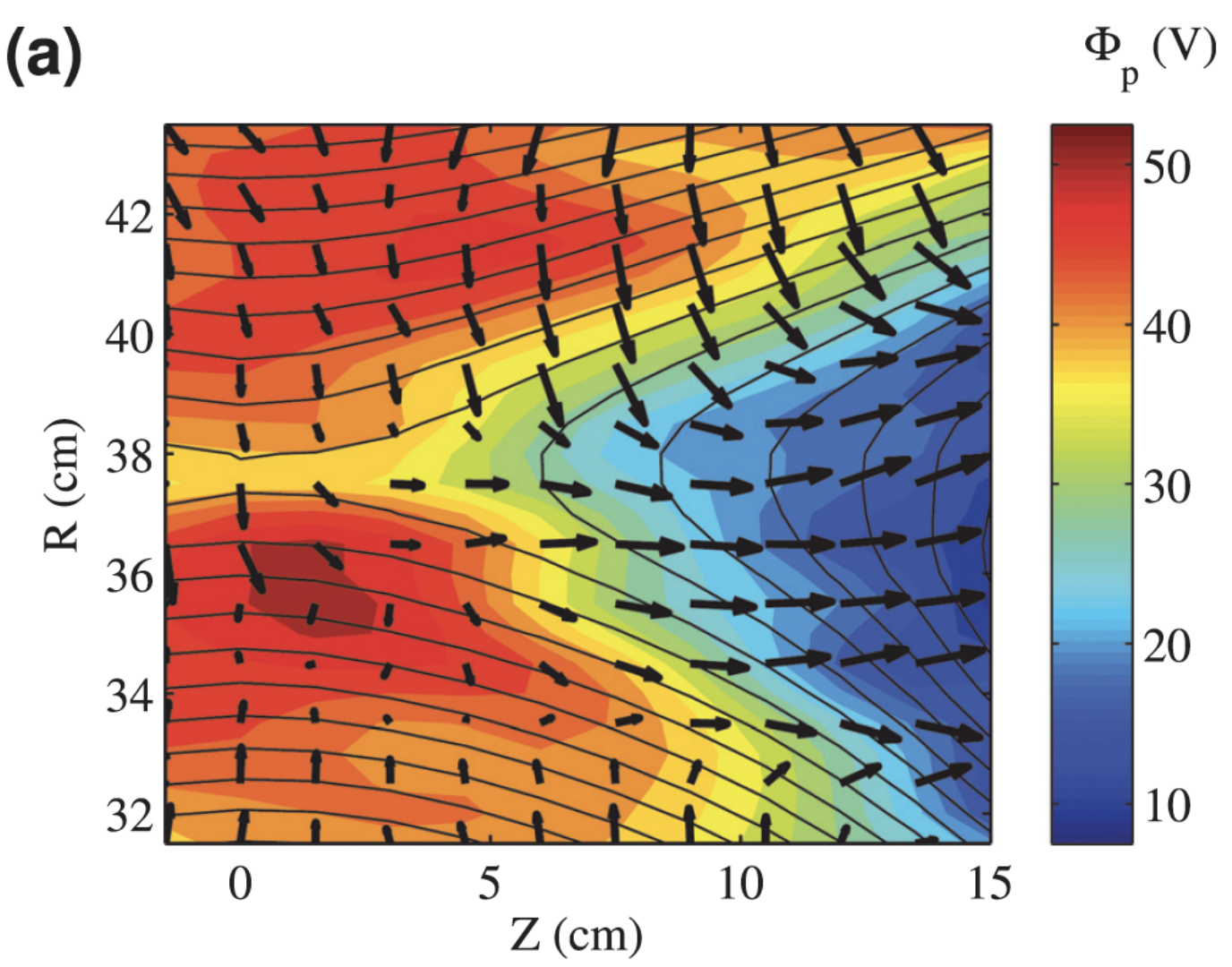}
    \includegraphics[width=5.3cm]{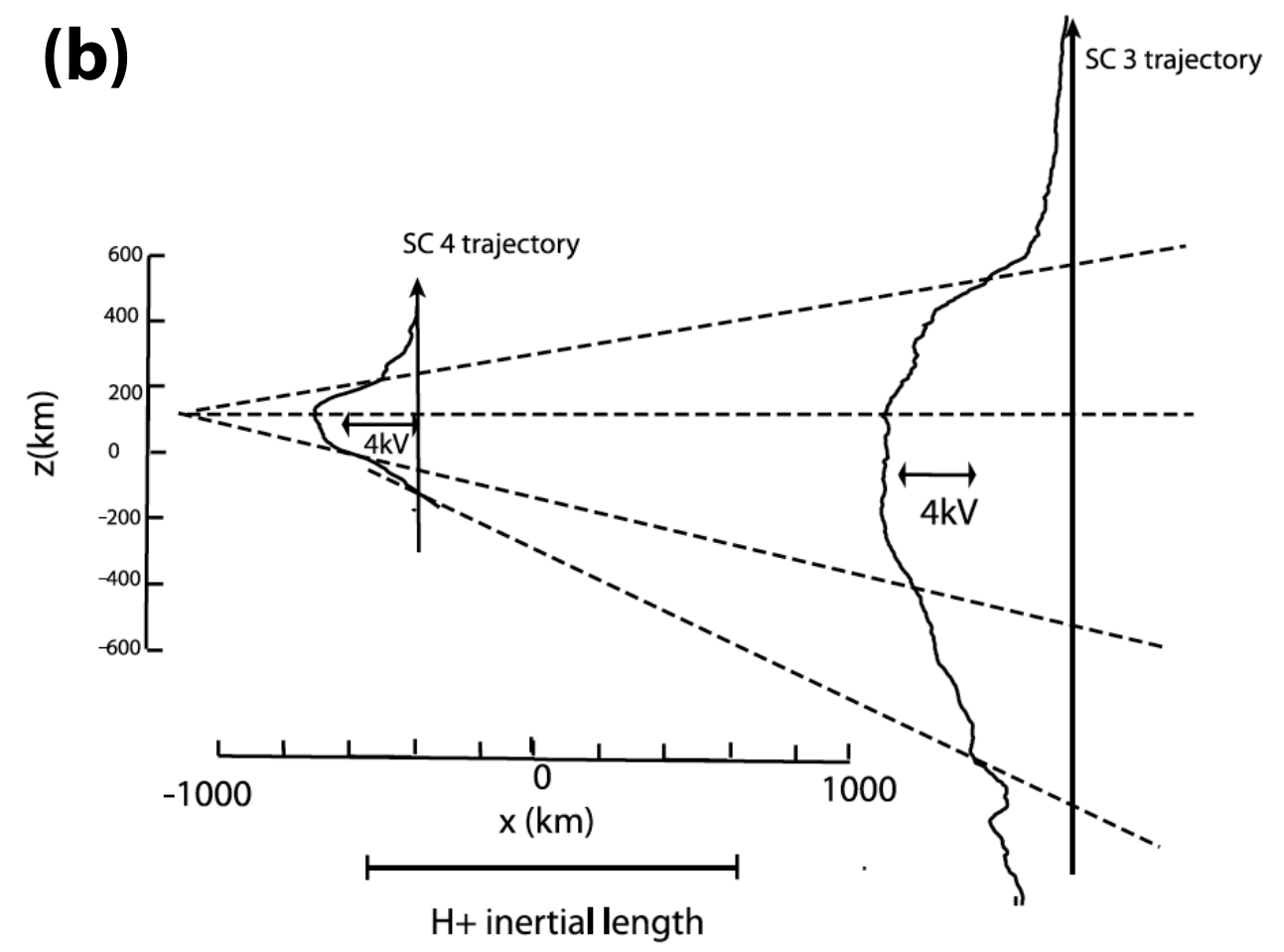}
    \caption{(a) Measured 2D Hall electric potential and ion in-plane flow in MRX during anti-parallel reconnection where a half of the saddle-type quadrupolar structure is shown. Adapted from \cite{yamada15}. (b) Measured Hall electric potential by two Cluster spacecraft during a magnetotail reconnection event, consistent with the expectation that the potential is deeper and wider more faraway from the X-line. Adapted from \cite{wygant05}.}
    \label{IDR1b}
\end{figure}

\subsubsection{Symmetric anti-parallel reconnection}

A textbook example measurement of the Hall magnetic and electric structures was by Polar spacecraft~\citep{mozer02} where a bipolar signature for both $B_y$ and $E_x$ was detected as the spacecraft traverses across current sheet on one of outflows during a rare event of symmetric, anti-parallel reconnection in Earth's magnetopause. Later with multiple spacecraft of Cluster, 2D structures of Hall magnetic and electric fields have been mapped statistically around X-line in Earth's magnetotail~\citep{eastwood10b}.

\begin{figure}[t]
	\centering
    \includegraphics[width=0.8\linewidth]{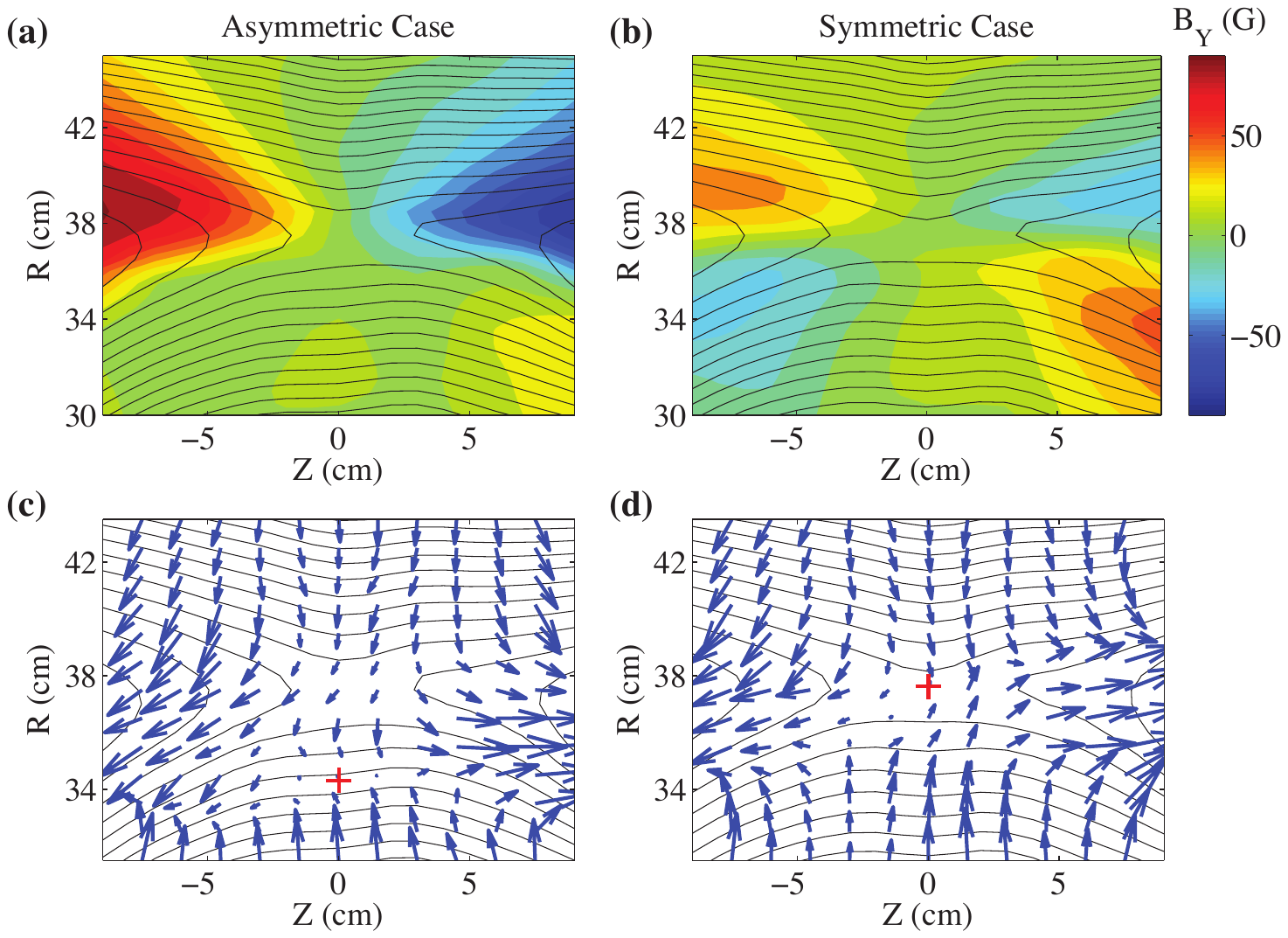}
	\caption{2-D profiles of the out-of-plane magnetic field ($B_y$) with contours of the poloidal flux for asymmetric (a) and symmetric (b) cases. Compared to the symmetric case, the quadrupole magnetic field component is enhanced on the high-density side ($R>37.5$ cm) and suppressed on the low-density side ($R<37.5$ cm). Black lines indicate contours of the poloidal magnetic flux which represent magnetic field lines. In-plane ion flow vector profiles for asymmetric (c) and symmetric (d) cases. For the asymmetric case, the ion inflow stagnation point is shifted to the low-density side. The upstream density ratio ($n_1/n_2$) for the asymmetric case is about 6, while it is about 1.2 for the symmetric case. Figure from \citet{yoo14}. }
	\label{fig:asym}
\end{figure}

Aiming to go beyond the 1D measurements by spacecraft, an effort was made in the laboratory experiments to directly capture instantaneous 2D quadrupolar structures in $B_y$ during anti-parallel reconnection. Figure~\ref{IDR1a}(a) and (b) show first such measurements from Magnetic Reconnection eXperiment or MRX~\citep{ren05} and Swarthmore Spheromak eXperiment or SSX~\citep{brown06}, respectively.  Furthermore, quantitative comparisons were made between MRX and 2D PIC simulations using corresponding parameters, showing excellent agreements on ion scales~\citep{ji08}, see Fig.\ref{IDR1a}(c). Since ions control the overall reconnection rate in collisionless reconnection~\citep{biskamp95,hesse99}, the convergence on the ion-scale kinetic structures between numerical prediction, laboratory experiment and space measurement essentially validated the concept of collisionless fast reconnection. In addition, since collisionality can be actively controlled in the laboratory, continuous transition has been demonstrated from slow Sweet-Parker collisional reconnection~\citep{ji98} without a significant $B_y$ structure to fast collisionless reconnection with a significant $B_y$ structure~\citep{yamada06}. The Hall electric field potential $\phi$ was also simultaneously measured by multiple spacecraft in the magnetotail on the ion scale at downstream~\citep{wygant05}, and on the electron scale across the current sheet~\citep{chen08b}. The structure is consistent with the 2D measurements in MRX where a half of the saddle-type quadrupolar potential structure is shown, see Fig.(\ref{IDR1b}).

\subsubsection{Asymmetric anti-parallel reconnection}

\begin{figure}[t]
	\centering
    \includegraphics[width=8cm]{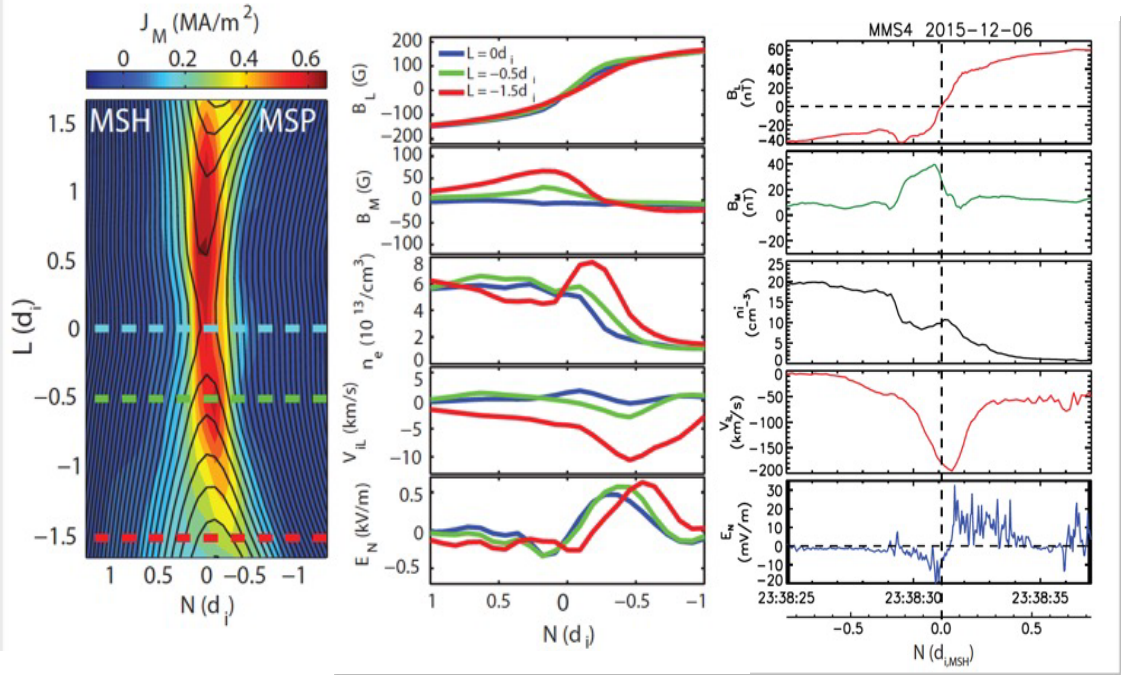}
	\caption{Comparisons of various profiles across asymmetric reconnection current sheet between MRX and MMS. (left panel) 2-D profiles reconnecting field lines and out-of-the-plane current density in MRX. (middle panel) Cross current sheet profiles of magnetic field, density, ion outflow and in-plane electric field at three different locations marked in the left panel. (right panel) Cross current sheet profiles of the same quantities during an magnetopause asymmetric reconnection event by MMS on December 6, 2015.}
	\label{MRX-MMS}
\end{figure}

Magnetic reconnection in nature often occurs with significant differences in the density, temperature, and magnetic field strength across the current sheet. A best example of this asymmetric reconnection is reconnection at the magnetopause~\citep{mozer11}, where the density ratio across the current sheet ranges from 10--100 and a magnetic field strength ratio of 2--4. 

In the laboratory, reconnection with a strong density asymmetry across the current sheet have been extensively studied and compared to space observations at the subsolar magnetopause~\citep{yoo14,yoo17,yamada18}. The ratio of the two upstream densities ranges from 5 to 10. It has been shown that strong density asymmetry alters the electric and magnetic field structures in diffusion regions. In IDR, the uniform reconnection electric field $E_y$ is approximately balanced by the Hall term $\bm{j}_\text{in-plane} \times \bm{B}/en$ on both upstreams. The asymmetry in density has to be compensated by asymmetry in $\bm{j}_\text{in-plane}$ since the in-plane magnetic fields are similar since the pressure balance is maintained by temperature asymmetry. The much larger $\bm{j}_\text{in-plane}$ significantly enlarge $B_y$ on the higher density side so that the quadrupolar structure becomes almost bipolar, as shown in Fig.~\ref{fig:asym}(a) and (b) \citep{yoo14}. In contrast, the in-plane electric field is much larger on the low density side since $\bm{E}_\text{in-plane} \approx j_yB/en$ where $j_y$ and $B$ are similar between two upstreams. As a result, the in-plane bipolar electrostatic field becomes almost unipolar~\citep{yoo17}. All these features agree with space observations~\citep[e.g.][]{mozer11,burch16}. Figure~\ref{MRX-MMS} show excellent agreements between MRX and an example of the MMS measurements at Earth's magnetopause on profiles of magnetic field components, density, ion outflow, and in-plane electric field.

\begin{figure}
	\centering
    \includegraphics[width=6cm]{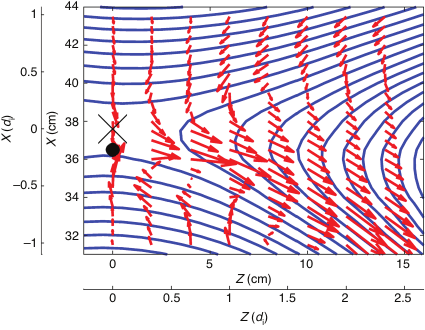}
	\caption{Electron dynamics observed during asymmetric reconnection in the MRX. In the reconnection plane, electron flows together with reconnecting field lines. The X marker at $(R, Z) = (37.6, 0)$ is the X-line and the black circle denotes the stagnation point of in-plane electron flow. Figure from \citet{yamada18}. }
	\label{IDR2}
\end{figure}

Strong density asymmetry also causes the shift of electron and ion stagnation points \citep{yoo14,yamada18}. The ion stagnation point is the location where the in-plane ion flow velocity vanishes. As shown in Fig.~\ref{fig:asym}(c) and (d), the ion stagnation point is shifted to the low-density side by about 3 cm ($\sim$0.5 $d_i$; $d_i$ is the ion skin depth) for the asymmetric case, while it is very close to the X-point for the symmetric case. 

The electron stagnation point is also shifted to the low-density side, as shown in the Fig.~\ref{IDR2}. The stagnation point denoted by the black dot is shifted by about 1 cm, which is about 0.15 $d_i$. These shifts are caused by the imbalance in the electron and ion inflow due to the density asymmetry. This overshooting of electrons from the magnetosheath (high-density) side is consistent with the well-known crescent-shape electron distribution function near the stagnation point~\citep{hesse14}, which is observed by MMS~\citep{burch16}. 

\begin{figure}[t]
	\centering
\includegraphics[width=0.8\textwidth]{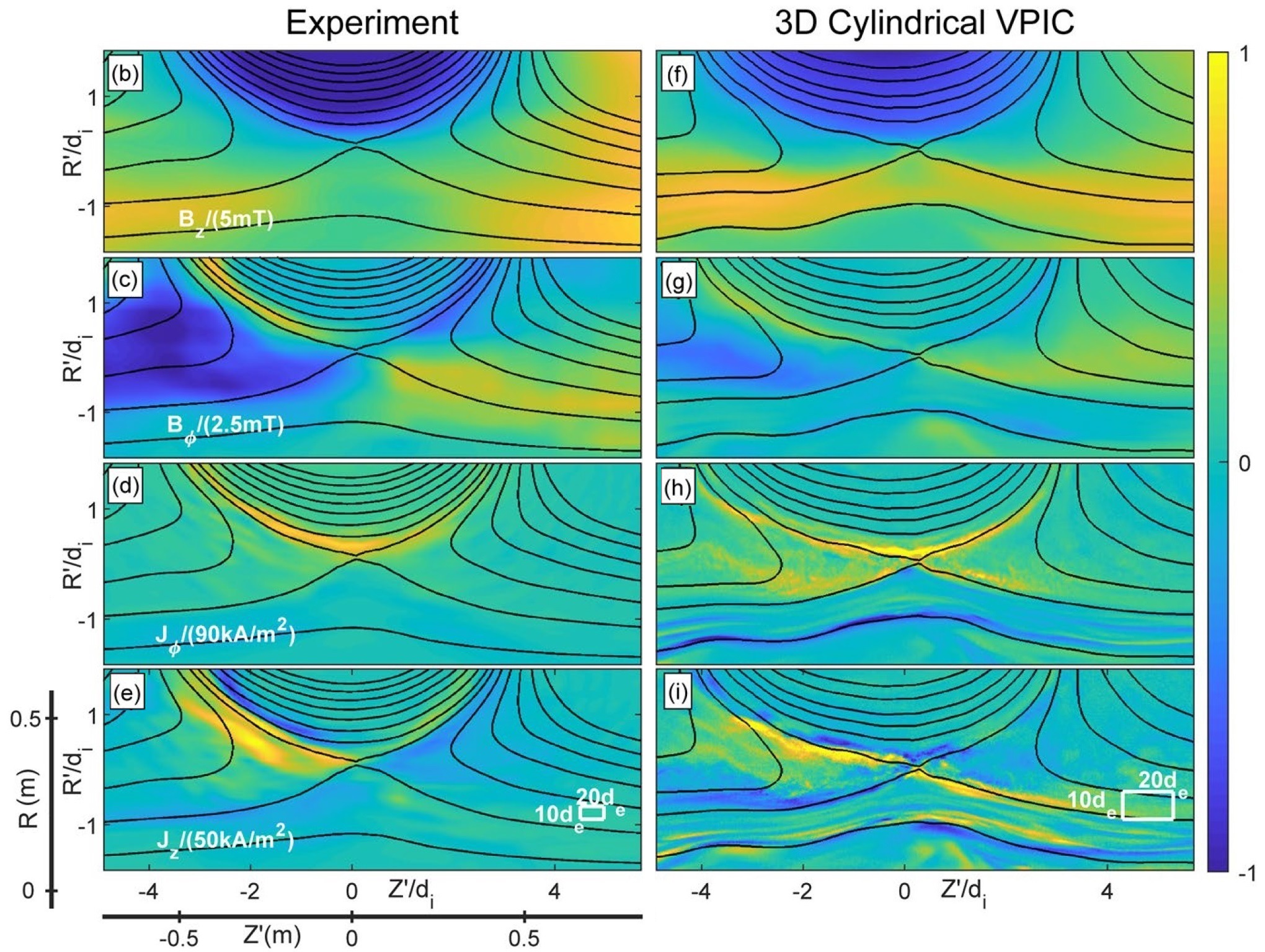}
\caption{(Panel b-e) Magnetic field and current components recorded in TREX during reconnection. (Panel f-i) Matched 3D kinetic simulation results reproducing the experimental results. After \cite{greess21}.}
\label{fig:fourCoil}
\end{figure}

The TREX experiment also explored asymmetric anti-parallel reconnection with the plasma density at large radii inflow being suppressed by a factor of about 4. Numerically, the TREX configuration was implemented in the cylindrical version of the VPIC code \citep{bowers09}, where properly scaled current sources increasing over time were added at the drive coil locations. Initial density and magnetic field profiles were set at the simulation based on experimental data.  As shown in Fig.~\ref{fig:fourCoil}, magnetic field and current structures similar to those of MRX are observed, and reproduced with remarkable agreement through matching numerical simulations \citep{olson21,greess21}.

\begin{figure*}[t]
\centering
\includegraphics[width=11cm, trim={1.5cm 0.3cm 1.2cm 0}, clip]{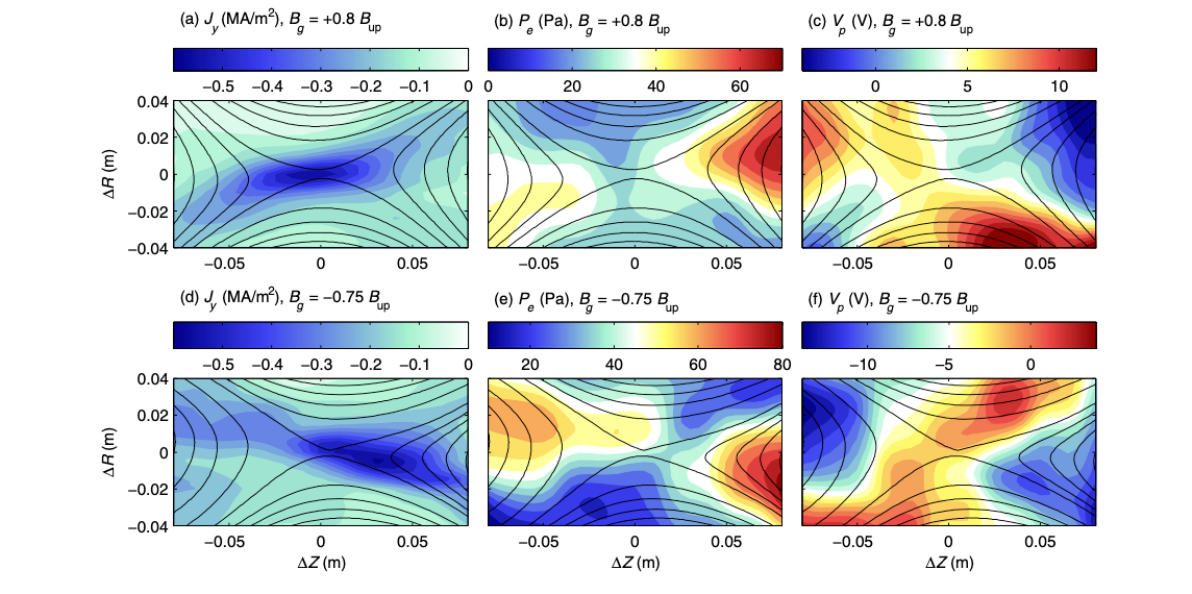}
\caption{2-D profile data showing observations of quadrupolar pressure variation during guide field magnetic reconnection. (a,d) Plasma current profile; (b,e) Plasma pressure; (c,f) Plasma potential.  Between (a-c) and (d-f) the sign of the guide field was reversed, leading to a change in the orientation of the quadrupolar profiles. After \cite{fox17}.}
\label{IDR3}
\end{figure*}

\subsection{IDR structures with a guide field}

Anti-parallel reconnection is a rather special magnetic geometry in nature where reconnection occurs often with a finite guide field, $B_g$. With the addition of $B_g$, the reconnecting field lines meet at an angle less than $180^\circ$, and a sufficiently strong guide field modifies the reconnection process by magnetizing the electrons and ions in the layer. The characteristic kinetic scale across the collisionless current sheet transitions from ion skin depth to ion sound Larmor radius ($\rho_s$) as $B_g$ increases.

A finite $B_g$ also introduces an in-plane electric field structure at the outer ideal scales even without a significant $V_y$. This is because in this case $E_\text{in-plane}=V_\text{in-plane} B_g$ where $V_\text{in-plane}$ is the in-plane flow due to reconnection. This $E_\text{in-plane}$ is required to satisfy ideal MHD condition $\bm{E} \cdot \bm{B}=E_yB_g + \bm{E}_\text{in-plane} \cdot \bm{B}=0$ as the reconnection electric field $E_y$ now has a parallel component which can extend over a large area. At upstream where the \textit{reconnecting} component $B_z$ dominates over the \textit{reconnected} component $B_x$, $E_z \approx -E_y (B_y/B_z) $ can even dominate the reconnection electric field $E_y$ under strong-guide field conditions.   Correspondingly, in the downstream where $B_z$ is small, $E_x \approx -E_y (B_y/B_x)$. As before, under quasi-steady conditions ($\partial B_y/\partial t \approx 0$) the in-plane electric field is well represented by a quadrupolar potential structure, $\bm{E}_\text{in-plane} = -\nabla \phi$. This potential structure, in turn, drives $\bm{E} \times \bm{B}$ drift for both electrons and ions to support the required in-plane, incompressible reconnection flow, $\bm{V}_\text{in-plane}$. This quadrupolar potential structure on the outer scales was observed in the VTF~\citep{egedal01,egedal03} with a strong guide field and shown to balance the global reconnection electric field in the upstream, as well as interactions with global MHD modes that drive reconnection~\citep{katz10}. However, this quadrupolar potential structure on the outer ideal scales has not been reported by space measurements.

This quadrupolar potential structure persists from the outer ideal scales to IDR with a characteristic scale of $\rho_s$ during guide field reconnection. When approaching $\rho_s$ scale, in addition to the \textit{incompressible} $\bm{u}_E=\bm{E}\times\bm{B}/B^2$ drift, the in-plane ion polarization drift, $\bm{u}_p=(m_i/eB^2)(\bm{u}_E\cdot \bm{\nabla})\bm{E}_\text{in-plane}$, becomes increasingly important. Here $m_i$ is ion mass. This cross-field ion polarization drift is \textit{compressible}, and it can generate density variation with electrons moving along the field line to satisfy quasineutrality~\citep{kleva95}. Combined with continuity equation, $(\bm{u}_E \cdot \bm{\nabla})n+n  \bm{\nabla} \cdot \bm{u}_p=0$, the predicted density variation follows $\ln{(n/n_0)} =  (m_i/eB^2)\nabla^2\phi$ with a quadrupolar structure. This density structure develops large electron pressure variations along the field lines until the third term on RHS of Eq.(\ref{GeneralizedOhm}) becomes important so that
\begin{equation}
E_\parallel=-\frac{\nabla_\parallel p_e}{en} \approx - \rho_s^2 \nabla_{\|}\nabla^2 \phi
\end{equation}
to reach a steady state in IDR. The quadrupolar density structure has been directly measured on MRX as shown in Fig.\ref{IDR3} during guide field reconnection. Such a structure was originally predicted from two-fluid extended MHD simulations \citep{aydemir92,kleva95}. \cite{oieroset16} have measured a plasma density variation consistent with such a quadrupolar structure during a current sheet crossing by MMS. The correspondence was observed in a symmetric guide-field reconnection event, and inferred through the comparison with simulations.  The crossing of the current sheet was sufficiently downstream that only a bipolar variation (half a quadrupole) was observed.

\subsection{EDR structures}\label{sec23}

The last two terms in Eq.(\ref{GeneralizedOhm}) are responsible in collisionless plasmas for magnetic field dissipation within electron diffusion region or EDR where electrons are demagnetized typically on the order of electron skin depth ($d_e$) or gyro radius ($\rho_e$). EDR is the location where field lines are finally reconnected from upstream toward downstream. In particular, the importance of off-diagonal terms in the electron pressure tensor in EDR has been predicted theoretically~\citep{vasyliunas75,lyons90}, demonstrated numerically~\citep{cai97,hesse99,pritchett01}, and explained physically~\citep{kulsrud05}. Unmagnetized electrons with an in-plane thermal speed $v_x$ or $v_z$ are subject to free acceleration by reconnection electric field $E_y$, generating large off-diagonal pressure $P_{xy}$ or $P_{zy}$, respectively, during their transit time in EDR. This manifests as spatial derivatives in the $y$ component of $\bm{\nabla}\cdot \bm{\Pi}_e$ in Eq.(\ref{GeneralizedOhm}). The competing alternative to this dissipation mechanism is the so-called anomalous resistivity based on 3D kinetic instabilities~\citep[][and references therein]{papadopoulos77} and they have been used numerically to reproduce the Petschek solution of fast reconnection~\citep{ugai77,sato79} since the early phase of reconnection research. There have been evidence from the MMS measurements for the laminar off-diagonal pressure tensor effect~\citep{torbert18,egedal18,egedal19} and also for possible importance of anomalous resistivity or 3D effect~\citep{torbert16,ergun17,cozzani21}.

\begin{figure}[t]
    \centering
    \includegraphics[width=6.5cm]{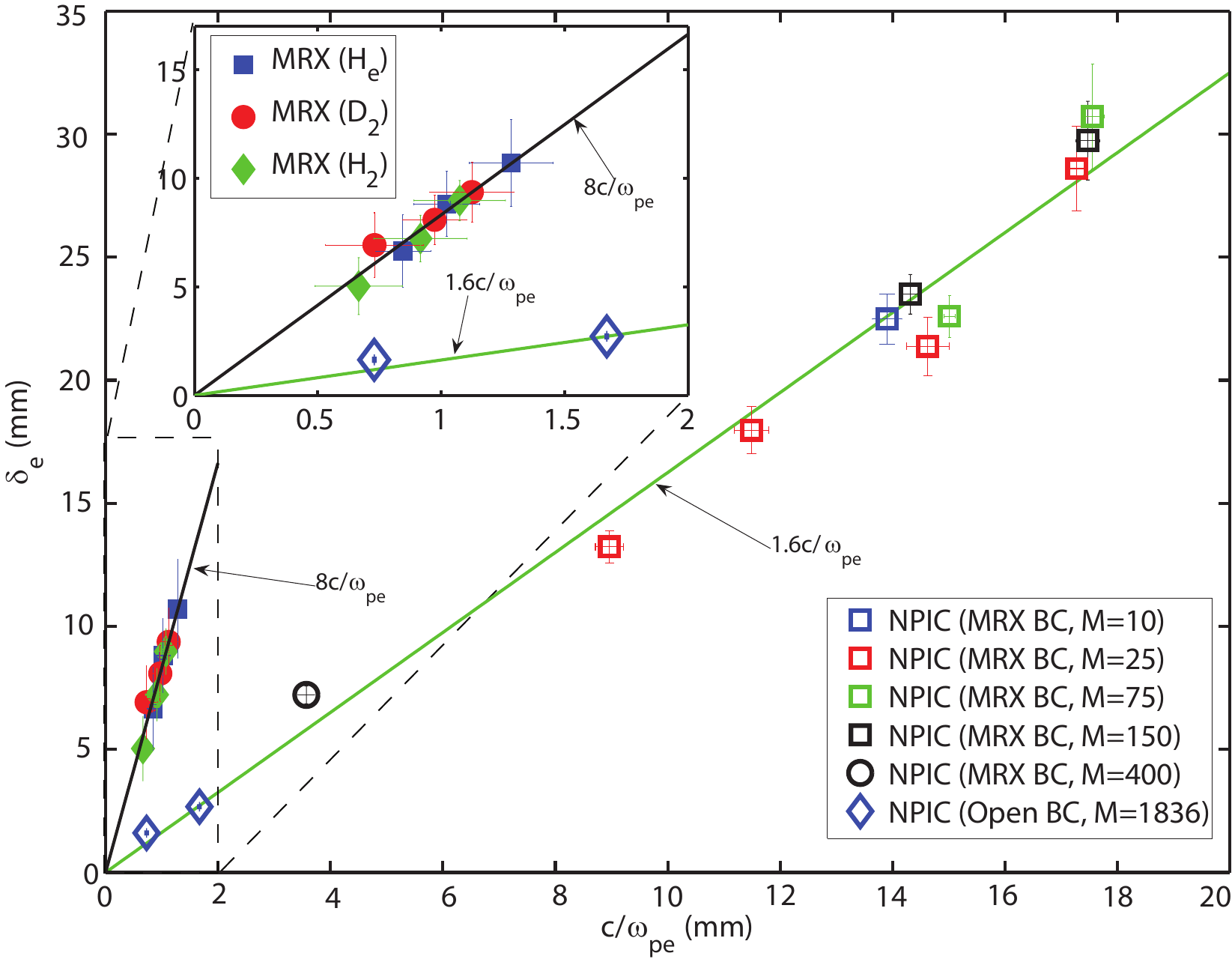}\includegraphics[width=5cm]{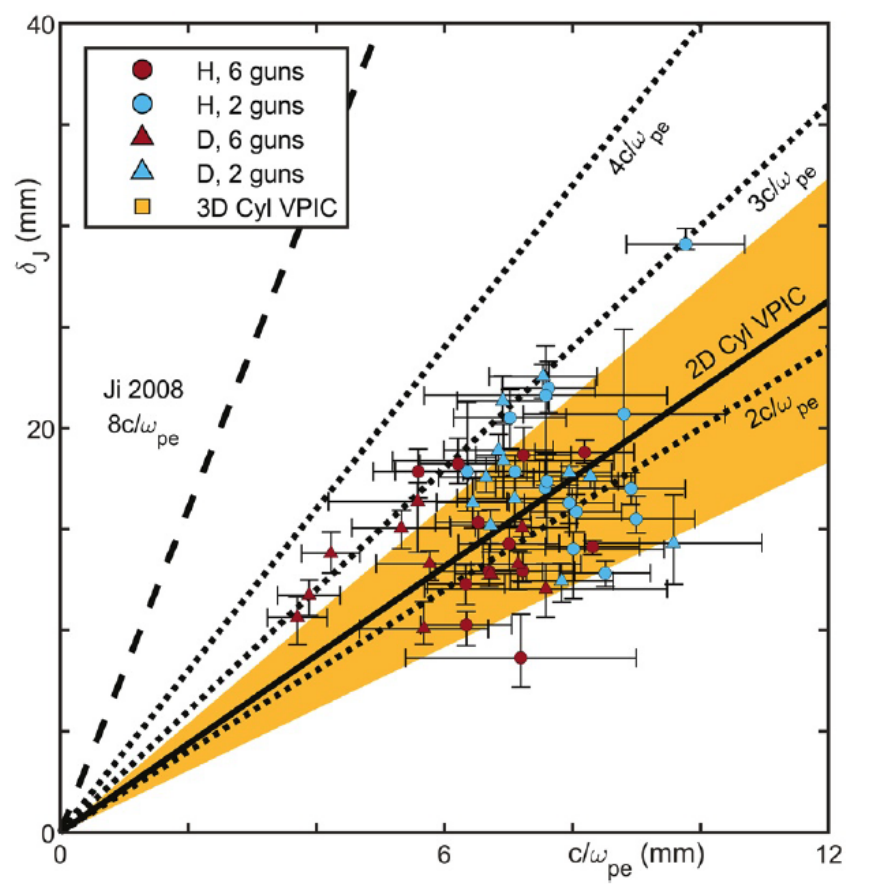}
    \caption{(a) Measured half width of EDR on MRX compared with 2D PIC simulations in cartesian geometry~\citep{ji08} (b) measured half width of EDR on TREX compared with 2D PIC simulation (solid line) and 3D (orange region) in cylinderical geometry~\citep{greess21}.}
    \label{EDR}
\end{figure}

The EDR has been also identified for the anti-parallel reconnection on MRX~\citep{ren08} as outgoing electron jets between two quadrants in the $B_y$ structure shown in Fig.\ref{IDR1a}(c). The importance of the off-diagonal pressure tensor in EDR is closely related to the magnitude and width of such electron jets~\citep{hesse99}. Compared with 2D PIC simulations in cartesian geometry, however, the electron jet speed is much slower and their layer half width is 3-5 times thicker~\citep{ji08}, as shown in Fig.\ref{EDR}(a). This discrepancy persisted even after incorporating finite collisions~\citep{roytershteyn10} and 3D effects via Lower Hybrid Drift Waves (LHDW, see later)~\citep{roytershteyn13} in the simulations when averaged over the $y$ direction. In contrast, the EDR has been recently studied on TREX and their measured half width agrees well with the predictions by 2D PIC simulations in cylinderical geometry~\citep{greess21}, shown in Fig.\ref{EDR}(b). 3D effects via LHDW can distort the EDR in the out-of-the-plane direction, weakly broadening the numerical directions of EDR width [orange region in Fig.\ref{EDR}(b)], but the off-diagonal pressure tensor effect remains dominant at each location. 

In addition to the differences in simulation geometries, there are several possibilities to resolve these different results. First, the anti-parallel reconnection in this comparison was driven symmetric on MRX (Fig.~\ref{IDR1a}) while asymmetric on TREX (Fig.~\ref{fig:fourCoil}). It is unclear whether symmetry plays a role in determining EDR thickness. Second, the colder ion temperature, $T_i \ll T_e$, at TREX may favor triggering LHDW which can distort the EDR~\citep{roytershteyn12}, compared with MRX where $T_i \sim T_e$. Third, there are also differences in measuring EDR: the ``jogging" method in which the EDR is rapidly swept over an 1D probe array in TREX may have higher effective spatial resolutions, requiring that the structures remain in same shape as confirmed experimentally~\citep{olson21}, while such a requirement is not needed for the 2D probe array but with less spatial resolutions on MRX. 

Furthermore, if there are sufficient scale separations between electron skin depth ($d_e$) and Debye length ($\lambda_D$) during anti-parallel reconnection, $d_e/\lambda_D=c/v_\text{th,e}>30$, the counter-streaming electron beams in the unmagnetized EDR are unstable to streaming instabilities~\citep{jara-almonte14}, possibly leading to efficient dissipation broadening EDR. Interestingly, this condition is equivalent to $T_e < 570$ eV which is generally satisfied in space, solar and laboratory plasmas, except in Earth's magnetotail and also in the typical PIC simulations where laminar anti-parallel reconnection is dominated by the electron pressure tensor effects~\citep[e.g.][]{torbert18,egedal19}. For guide field reconnection, this condition should be revised to $\rho_e/\lambda_D = \omega_{pe}/\omega_{ce}= (\sqrt{\beta_e/2}) d_e/\lambda_D>30$ implying the importance of electron beta, $\beta_e$. Obviously, further research is needed to resolve these differences in order to understand better when and how 2D laminar or 3D anomalous effects dominate the dissipation in EDR.

\section{Energy conversion and partition}

\subsection{Magnetic energy dissipation at the X-point}

\begin{figure}[t]
	\centering
    \includegraphics[width=0.8\linewidth]{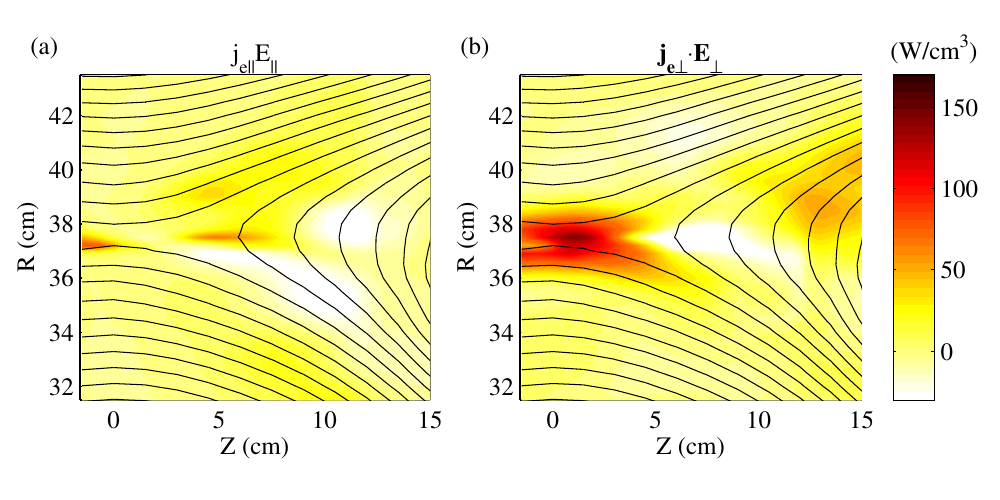}
	\caption{Comparison of two compositions of energy deposition rate measured in MRX for symmetric, anti-parallel magnetic reconnection; (a) $j_{e\parallel}E_{\parallel}$ and (b) $\bm{j}_{e\perp}\cdot\bm{E}_{\perp}$. Figure from \citet{yamada16}.  }
	\label{fig:sym_je}
\end{figure}

The primary consequence of magnetic reconnection is the impulsive dissipation of excessive free energy in magnetic field to plasma charged particles. The energy dissipation near the X-point (inside the EDR) is dominated by electron dynamics, as the electron current is much stronger than the ion current in the EDR. The rate of the energy conversion from the magnetic to plasma kinetic energy per unit volume can be quantified by $\bm{j}\cdot\bm{E}$ in the laboratory frame. This frame-dependent $\bm{j}\cdot\bm{E}$, however, is not much different from the frame-independent quantity of $\bm{j}\cdot\bm{E}^{\prime}$ where $\bm{E}^\prime = \bm{E}+\bm{V}_e \times \bm{B}$ \citep{zenitani11} in the EDR especially near the X-point, since electrons are unmagnetized. Thus, we will only discuss the quantity of $\bm{j}\cdot\bm{E}$ here for simplicity. 

During anti-parallel reconnection, magnetic energy dissipation near the X-point is dominated by the perpendicular component of $\bm{j}_e\cdot\bm{E}$, $\bm{j}_{e\perp}\cdot\bm{E}_{\perp}$, in both symmetric~\citep{yamada14,yamada16} and asymmetric cases~\citep{yoo17,yamada18}. Figure \ref{fig:sym_je} shows a clear dominance of $\bm{j}_{e\perp}\cdot\bm{E}_{\perp}$ (panel b) over $j_{e\parallel}E_{\parallel}$ (panel a) near the X-point at $(R,Z) = (37.5, 0)$ cm during symmetric, anti-parallel reconnection in MRX. This agrees well with space where $\bm{j}_{e\perp}\cdot\bm{E}_{\perp}$ is strongest near the stagnation point \citep{burch16,yamada18}. Furthermore, the perpendicular electric field near the X-point is dominated by the out-of-the-plane reconnection electric field which can directly accelerate electrons~\citep{zenitani01} as shown during an magnetotail reconnection event measured by MMS~\citep{torbert18}, and also recently during anti-parallel reconnection driven by lasers~\citep{chien23} where an accelerated electron beam was detected. 

\begin{figure}[t]
	\centering
 	\includegraphics[width=0.5\linewidth]{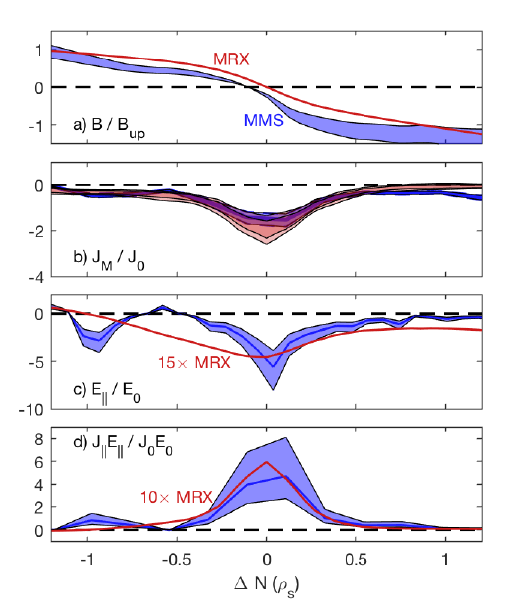}
	\caption{
	Scaled comparison of MRX (red curves and bands) and MMS (blue bands) data from the event of \cite{eriksson16}, for cuts of the reconnecting magnetic field (a), out-of-the plane current density (b) parallel electric field (c), and the parallel component of energy dissipation rate (d) from \cite{fox18}.
	}
	\label{fox18fig2}
\end{figure}

If there is a significant guide field, however, the energy conversion is dominated by the parallel component, $j_{e\parallel}E_{\parallel}$~\citep{fox18,pucci18,bose22}, consistent with the MMS observation~\citep{wilder18}. This difference is mainly related to the fact that the energy conversion inside the EDR is mostly through the out-of-plane reconnection electric field. Without a guide field, the reconnection electric field is mostly perpendicular to the magnetic field, while it becomes mostly parallel to the magnetic field with a sizable guide field. Figure~\ref{fox18fig2} shows direct and scaled comparisons between MRX with a guide field of about 0.6 times of the reconnecting field~\citep{fox17} and MMS data with a guide field of about 3.5 times of the reconnecting field~\citep{eriksson16}. When normalized properly, profiles of magnetic field and current density agree with each other within error bars. A similar conclusion was obtained when compared with another MMS event with lower guide field~\citep{wilder18}. In both cases $\bm{j} \cdot \bm{E}$ in the current sheet is dominated by $j_\parallel E_\parallel$, consistent with numerical predictions~\citep{pucci18}. The peak values of parallel electric field, however, is larger by an order of magnitude in MMS than in MRX. This highlights the importance in our further understanding energy conversion by reconnection~\citep{ergun16a}, including questions on where do these intense parallel electric fields come from and what effects do they have on plasma heating and acceleration.

\subsection{Energy conversion}

Particle heating and acceleration local to the reconnection region have been directly measured in details in the laboratory~\citep{hsu00,brown02,stark05,ono11,tanabe15,yoo13,yoo14}. During anti-parallel reconnection whether symmetric or asymmetric on MRX, incoming ions from upstream are directly accelerated by the in-plane electrostatic electric field, $\bm{E}_\text{in-plane}$, in the IDR~\citep{yoo13,yoo14} (see Fig.\ref{IDR1b}(b)) before they are ``remagnetized" at further downstream converting flow energy to thermal energy. Although $\bm{E}_\text{in-plane} \approx -(\bm{V_e} \times \bm {B})_\text{in-plane}$ is non-dissipative for electrons within the IDR (but outside EDR), it can energize ions via $en \bm{V}_i \cdot (\bm{V_e} \times \bm {B})_\text{in-plane}$~\citep{liu22}. This has been confirmed numerically~\citep{yoo14b,yamada18}.

During strong guide field reconnection in VTF, ion heating was observed and interpreted~\citep{stark05} as magnetic moment conservation was broken due to strong motional variation of the in-plane electric field~\citep{egedal03}, $(\bm{v} \cdot \bm{\nabla}) \bm{E}_\text{in-plane}$. A key dimensionless parameter $e \nabla^2 \phi / m_i B^2 \gtrsim 1$ was identified to demagnetize and energize ions~\citep{stark05}. Ions are heated downstream of magnetic reconnection during plasma merging with a significant guide field~\citep{ono11}.

Electron heating is mostly localized to the EDR near the X-line during symmetric anti-parallel reconnection as implied by the large value of $\bm{j}\cdot \bm{E}$ there~\citep{yoo14b} or along the low-density side of separatrices during asymmetric anti-parallel reconnection on MRX~\citep{yoo17}. While parallel electric field is expected to explain a large fraction of electron temperature increase~\citep{egedal13,yoo17}, other mechanisms, such as by various wave activities (see below), are not excluded~\citep{ji04,zhang23}. Electron heating is also measured during guide field reconnection in the electron-only region~\citep{shi22} and in the electron-ion region on MRX~\citep{bose22}. Strong electron heating was observed within the current sheet during plasma merging~\citep{tanabe15}. These results are in general agreement with MMS results on significant electron energization within EDR~\citep{eastwood20}.

Direct measurements of particle acceleration local to the reconnection region are generally difficult in the laboratory, despite many acceleration mechanisms have been proposed and studied intensively numerically~\citep{ji22}. They include direct acceleration by reconnection electric field~\citep{zenitani01}, parallel electric field~\citep{egedal13}, Fermi acceleration~\citep{drake06}, and betatron acceleration~\citep{hoshino01}. Accelerated electrons along magnetic field were measured by an energy analyzer~\citep{gekelman85} during reconnection although in a different region. On VTF where reconnection is driven dynamically with a strong guide field, a population of energized tail of electrons along the field line were detected to increase by factors of several, doubling an effective temperature from $\sim20$ eV to up to 40eV~\citep{fox10,fox12a}. Electron jets at electron Alfv\'en speed have been directly detected by Thomson scattering diagnostics during guide field electron-only reconnection~\citep{shi22}. More recently, non-thermal electrons with energies of $\sim 100T_e$ due to reconnection electric field of anti-parallel reconnection at low-$\beta$ driven by lasers were directly detected with an angular spread consistent with simulation~\citep{chien23}. The later supports an astrophysical conjecture to accelerate electrons by reconnection to high energies beyond the synchrotron burnoff limit~\citep{cerutti13}.

\subsection{Energy partition}

One of the advantages of laboratory experiments over the space measurements is that 2D profiles of key plasma and field parameters can be obtained by repeating measurements over a similar set of discharges. These 2D profiles can be used for a quantitative study of energy conversion and partition inside the IDR on MRX~\citep{yamada14,yoo17,bose22}, where the method of the energy inventory analysis has been explained in detail. The incoming magnetic energy, for example, can be obtained by integrating the corresponding Poynting flux ($E_yB_z/\mu_0$) at the boundary surface. The electron (ion) energy gain can be obtained by integrating $\bm{j}_e\cdot\bm{E}$ ($\bm{j}_i\cdot\bm{E}$) over the entire volume of the analysis. 

\begin{table}[!ht]
\caption{Summary of the energy inventory studied in the laboratory for three cases and their counterparts based on PIC simulations for two cases~\citep{yamada14,yoo17,yamada18,bose22}. Typical errors for these numbers are about 10--20\%. The guide field was about 0.7 times of reconnecting field for the guide field reconnection case. One study of space data for a symmetric antiparallel case in Earth's magnetotail~\citep{eastwood13} is also listed despite of the large uncertainties in determining incoming magnetic energy and sizes of the volume~\citep{yamada15}.}
\vspace{3mm}
\centering
\begin{tabular}{|l|l|r|r|r|}
\hline
Case & Incoming (MW) & Outgoing &  Electron & Ion  \\ \hline \hline
Symmetric, antiparallel, lab & 1 (1.9 $\pm$ 0.2) & 0.45 & 0.20 & 0.35 \\ \hline
Symmetric, antiparallel, PIC & 1			& 0.42 & 0.22 & 0.34 \\ \hline
Symmetric, antiparallel, space & 1			& 0.1-0.3  & 0.18 & 0.39\\ \hline\hline
Asymmetric, antiparallel, lab & 1 (1.4 $\pm$ 0.2) & 0.44 & 0.25 & 0.31 \\ \hline
Asymmetric, antiparallel, PIC & 1			& 0.43 & 0.25 & 0.32 \\ \hline\hline
Symmetric, guide field, lab & 1 (1.5 $\pm$ 0.2) & 0.65  & 0.15 & 0.29\\ \hline
\end{tabular}
\label{table:inventory}
\end{table}

Table \ref{table:inventory} summarizes the energy partition for three cases in the lab, two cases of numerical simulations, and one case from the space measurements. In all cases, the ion energy gain exceeds that of electrons. Compared to antiparallel reconnection, the total energy conversion is less effective for the case with a guide field at a strength comparable to the reconnecting field component. In all cases, both electron and ion energy gain is dominated by increase in the thermal energy; the flow energy increase is negligible especially for electrons. These results are in general agreement with space observations~\citep{eastwood13} which is also listed in the table for comparisons, despite that they carry large uncertainties due to limited available data. Nonetheless, the fact that all these numbers agree with each other in the ballpark suggests that energy conversion and partition in locations near the X-line during collisionless reconnection are reasonably quantified.

\section{Plasma waves}

While magnetic reconnection converts magnetic energy to plasma, various free energy sources for waves and instabilities are available especially in or near the diffusion regions and separatrices, in the form of spatial inhomogenuity, relative drift between ions and electrons (or electric current), or kinetic structures in particles' velocity distribution functions. This section reviews relevant studies of generated plasma waves in the vicinity of diffusion regions of collisionless reconnection in the laboratory in comparisons with space measurements.

\subsection{Whistler waves}

\begin{figure}
	\centering
    \includegraphics[width=0.8\linewidth, trim={4mm 8mm 0 0}, clip]{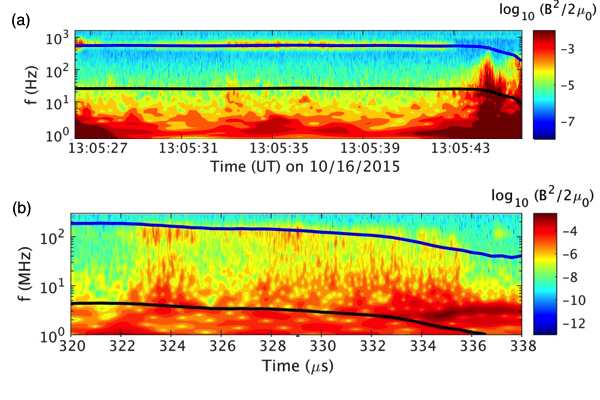}
	\caption{Comparison of the whistler wave activity during asymmetric reconnection observed in space (a) and MRX (b). Blue lines indicate the half of the local electron cyclotron frequency ($f_{\text{ce}}$), while black lines indicate the local lower hybrid frequency ($f_{\text{LH}}$). Near the separatrix on the low-density side, whistler waves near 0.5 $f_{\text{ce}}$ are observed. After \cite{yoo18}.  }
	\label{fig:whistler}
\end{figure}

One of these waves is whistler waves that can be generated by either electron beams or temperature anisotropy as summarized by \cite{khotyaintsev19}. During asymmetric reconnection, the separatrix region on the low-density (magnetospheric) side is unstable to the lower hybrid drift waves (LHDW)~\citep[][see below]{krall71} due to the large density gradient across magnetic field. This instability enhances the electron transport and heating near the separatrix region~\citep{le17}. In this region, electrons with a high parallel velocity can be quickly transported to the exhaust region along the turbulent field lines due to LHDW, leaving behind a population of electrons with temperature anisotropy due to a tail with higher perpendicular energy. This temperature anisotropy generates whistler waves around $0.5 f_{\text{ce}}$ near the separtrix on the low-density side~\citep{yoo18,yoo19}. 

Figure \ref{fig:whistler} shows this anisotropy-driven whistler wave observed by MMS (a) and in MRX (b). The color contour shows the energy in fluctuations in the magnetic field. Clear whistler wave activity around the half of the local electron cyclotron frequency ($0.5f_{\text{ce}}$), which is indicated by blue solid lines, is observed in both space and laboratory. In both cases, the measurement location was initially just outside of the separatrix region and moved to the exhaust region around 13:05:43 for the panel (a) and 334 $\mu$s for the panel (b). Broad fluctuations mostly below the local lower hybrid frequency ($f_{\text{LH}}$, denoted by black lines) also exist in both measurements. Note that LHDW-driven fluctuations are strongest just before the measurement location enters into the exhaust region. It should be also noted that the whistler wave activity disappears in the exhaust region. LHDW will be discussed below.

\subsection{Electrostatic waves}

A variety of electrostatic high-frequency waves have also been observed in the laboratory during reconnection events. Above $f_\text{LH}$, these waves have multiple names, including R-waves [after the $R=0$ branch in the Clemmow-Mullaly-Allis (CMA) diagram~\citep{stix92}], electrostatic whistlers, or Trivelpiece-Gould modes [from early laboratory contexts~\citep{trivelpiece59}]. These waves extend from $\sim f_{LH}$ to $\min(f_{pe}, f_{ce})$.  Under most laboratory as well as space conditions, $f_{ce} < f_{pe}$, so the waves exist up to $f_{ce}$.  For the waves to be electrostatic, $k d_e > 1$, where $k$ is the wavenumber and $d_e$ the electron skin depth. The electrostatic branch has the dispersion relation $\omega = \omega_{ce} k_\parallel/k$, which allows a broadband collection of waves with parallel phase velocities $\omega/k_\parallel$ resonant with super-thermal electron populations. At longer wavelength, when $k d_e < 1$, these waves transition to the classical electromagnetic whistlers ($\omega = \omega_{ce} d_e^2 k_\parallel k$). At lower frequencies $f \sim f_\text{LH}$, the waves increasingly interact with the ions. In those cases, the perpendicular group velocity of waves becomes very small, so that wave packets can stay localized to regions with energized electrons for efficient growth. Theory predicts that there are multiple sources of free energy which can drive the waves, including beam resonance (inverse Landau damping), gyro-resonance driven by $T_\parallel > T_\perp$, or gradients in density, temperature, or in fast electron components~\citep{fox10}. Most interestingly, the waves driven by gradients lead to maximum growth in the lower-hybrid range frequencies ($f \sim f_{LH}$), and are related to quasi-electrostatic lower-hybrid drift waves (see below).

\cite{gekelman85} also reported the detection of these waves on LAPD and suggested that they are generated by the measured energetic electron tail in the 3D velocity space, either by anisotropy mechanisms or inverse Landau damping. High-frequency electrostatic waves have been also detected on VTF only when guide field reconnection is strongly driven~\citep{fox10}. This was consistent with a picture where the reconnection events would drive energetic electrons, which in turn would drive waves. The parallel phase speed was observed to be resonant with superthermal electrons, $\omega / k_\parallel > v_{te}$. The spectrum typically consisted of a broad spectrum from near $f_\text{LH}$ and extending to a very clear cutoff at $f_{ce}$~\citep{fox10}. 

\begin{figure}
\centering
\includegraphics[width=3in]{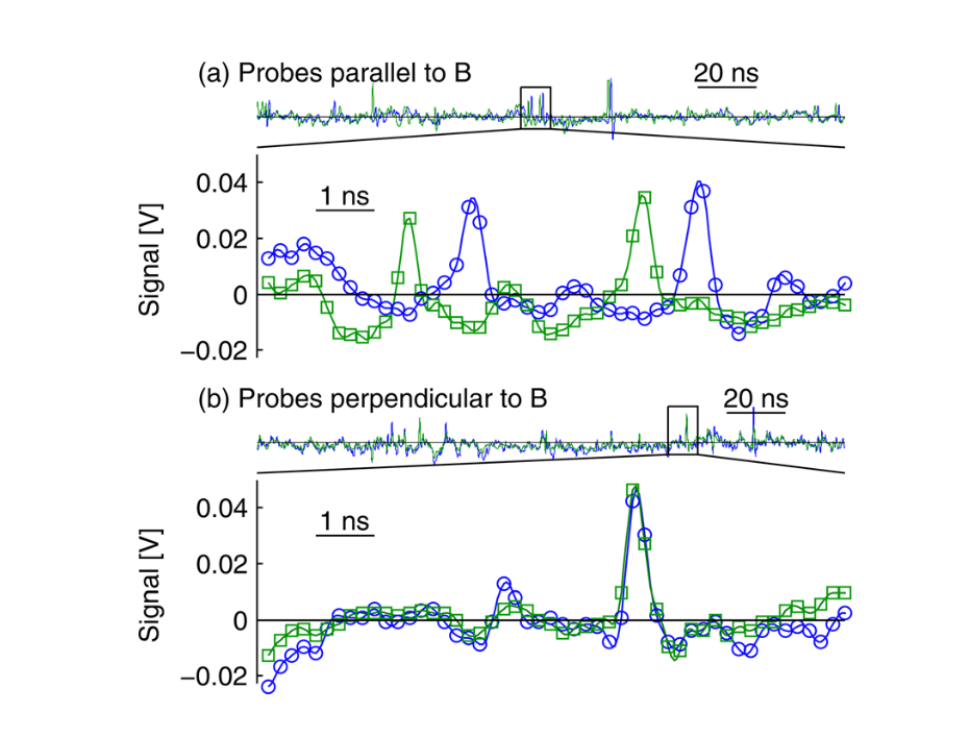}
\caption{Observation of phase-space-holes electrostatic structures driven during magnetic reocnnection events. a)Propagation between two closely-space probes parallel to the magnetic field, b) simultaneous observation on two probes oriented perpendicular to the magnetic field.  The time delays combined with known probe separate give the typical size and velocity of the electron holes, which is superthermal compared to the electron temperature.  From \cite{fox12a}.}
\label{FigFox2012}
\end{figure}

Given strong beam components, electrostatic waves can often be driven to very large amplitude, which can lead to the formation of non-linear wave structures.  One mechanism is that the waves can grow to large amplitude and trap resonant electrons. This leads to a so-called ``electron phase-space hole'' structure, also called a Bernstein-Greene-Kruskal (BGK) solitary structures~\citep{bernstein57}, or electrostatic solitary waves (ESW). The latter has been observed in many places in space and as well as during reconnection events in magnetopause~\citep{matsumoto03} and magnetotail~\citep{cattell05}, and summarized recently by \cite{khotyaintsev19}. These electron phase space holes were directly observed on VTF~\citep{fox08,fox12a} and indicate that the strong electric fields in the reconnection region pull-out strong beam components of the electron population, exciting these hole structures. Electron holes have also been directly generated in electron-beam experiments~\citep{lefebvre10}. Figure~\ref{FigFox2012} shows observations of electron hole phenomena during the strong wave turbulence during VTF reconnection events.  The structures are positive potential ($\phi > 0$) which is consistent with electron trapping. More recently, ESW or electron space holes are observed during guide field reconnection within the diffusion region~\citep{khotyaintsev20} or in the separatrix~\citep{ahmadi22} in the magnetopause where they may play an important role in electron heating.

There is a renewed interest in ion acoustic wave (IAW)~\cite[][and references therein]{papadopoulos77}, which is an unmagnetized short-wavelength electrostatic wave. The IAW can be driven unstable by relative drift between ions and electrons or equivalently electric current which is expected to be intense around the X-line. Anomalous resistivity based on the IAW-like waves has been used to numerically generate Petschek solution fast reconnection since \cite{ugai77,sato79}. Despite pioneering laboratory detection during a relatively collisional reconnection~\citep{gekelman84}, however, the importance of IAWs for reconnection has been quickly dismissed due to the widely observed high ion temperature $T_i \sim ZT_e$ which is known to stabilize IAW via strong ion Landau damping. Only in a very recent laboratory experiment using lasers~\citep{zhang23}, strong IAW bursts and the associated electron acoustic wave (EAW) bursts were detected by collective Thomson scattering in the exhaust of anti-parallel reconnection where $T_i \ll ZT_e$ due to high $Z (\sim 18)$ of ions. These IAW and EAW burst were successfully reproduced by PIC simulations showing that strong IAWs generate a double layer, which induces electron two-stream instabilities leading to EAW bursts and electron heating as observed experimentally. These new experimental results are consistent with recent space observations~\citep{uchino17,steinvall21} which detected IAWs during reconnection when sufficient cold ions are present, and may be relevant to the outstanding questions on large parallel electric field measured by MMS~\citep{ergun16b}. These new results also raised a legitimate question on whether the high ion temperature is a universal observation and thus whether IAW should be dismissed as an anomalous dissipation mechanism in collisionless plasmas. In fact, recent detection of monochromatic IAWs and associated electron heating in solar wind when ions are cold~\citep{mozer22} speaks for the needs to revisit this topic, as direct measurements of ion temperature are rare for solar and astrophysical plasmas in general.

\subsection{Lower hybrid drift waves and current sheet kinking}

Lower hybrid drift waves (LHDW) have been a candidate for anomalous resistivity and transport in the diffusion region due to its ability to interact with both electrons and ions. The free energy source of LHDW is the perpendicular current to the magnetic field~\citep{davidson75}. Depending on the local plasma and field parameters, LHDWs may be either quasi-electrostatic (ES-LHDW)~\citep{carter01,hu21} or electromagnetic (EM-LHDW)~\citep{ji04,yoo14}. With a similar electron temperature and perpendicular current, plasma beta ($\beta$) is the key parameter to determine the type of waves; for low $\beta$ (typically below unity), the ES-LHDW mode propagating nearly perpendicular to the local magnetic field is unstable, while EM-LHDW mode propagating obliquely to the magnetic field is excited when $\beta$ is high \citep{yoo20}. 

\begin{figure}[t]
	\centering
    \includegraphics[width=0.5\linewidth]{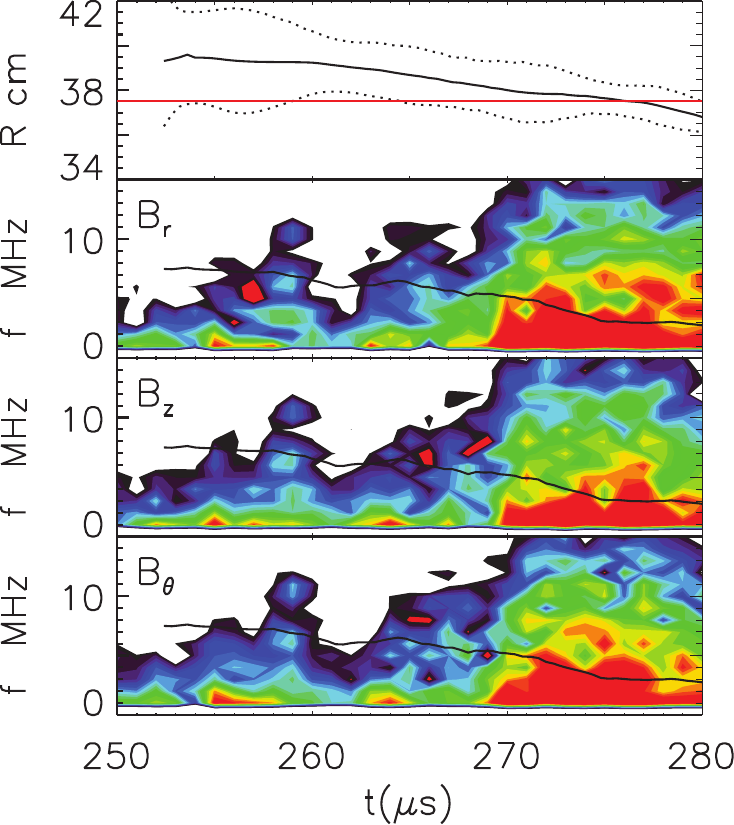}
	\caption{Detection of electromagnetic lower-hybrid drift waves in the current sheet center during anti-parallel reconnection on MRX. Wave powers are color coded (red high and white low) in spectrograms where lower hybrid frequency is indicated by black line using upstream reconnecting field. Top panel shows location of the probe (red) and the current sheet (center as black solid line and edges as dashed lines). When the current sheet center moves close to the probe, high-frequency magnetic fluctuations are detected. Figure from \citet{ji04}.  }
	\label{fig:EM}
\end{figure}

During the anti-parallel reconnection, plasma $\beta$ varies rapidly in the current sheet. At the current sheet edge where $\beta$ is low, the ES-LHDW mode has been observed~\citep{carter01,yoo20} consistent with theoretical expectation~\citep{daughton03} and space observation by Polar spacecraft~\citep{bale02}. The obliquely propagating EM-LHDW mode has been observed in the current sheet center where plasma $\beta$ is high and electric current is large~\citep{ji04,yoo14}, as well as in the immediate downstream~\citep{renthesis07}. An example is shown in Fig.~\ref{fig:EM} from MRX where large-amplitude electromagnetic waves were detected when the current sheet center moves close to the probe during anti-parallel reconnection~\citep{ji04}, consistent with numerical simulations~\citep{daughton04}. Both ES-LHDW and obliquely propagating EM-LHDW have been also observed by Cluster spacecraft in a thin current sheet in magnetotail~\citep{zhou09} and recently by MMS in magnetopause~\citep{ergun17}. More recent measurements on MRX show the EM-LHDW becomes increasingly organized with larger amplitude with guide field~\citep{stechow18}. For more measurements of LHDW in and around diffusion regions in space with varying influence on anomalous resistivity and viscosity, see recent reviews by~\cite{khotyaintsev19} and~\cite{graham23}.

Many of the observed wave characteristics of EM-LHDW, such as propagation direction and polarization, have been qualitatively explained by a local two-fluid theory~\citep{ji05} as an instability caused by reactive coupling between the backward propagating whistler wave and the forward propagating sound wave when the relative drifts between electrons and ions are large. The wave amplitude has been observed to correlate positively with fast reconnection~\citep{ji04}, consistent with quasilinear theory on their possible importance for anomalous resisitivity~\citep{kulsrud05}. The waves have been also reproduced in 3D PIC simulations performed in MRX geometry in a cartesian coordinate, but they failed to explain the observed broadened width of EDR~\citep{roytershteyn13}. Possible solutions to this discrepancy include differences in the simulation geometry and parameters, as well as measurement resolutions as discussed in Sec.\ref{sec23}. It is noted that the current sheet kinking that was observed on TREX and associated simulations~\citep{greess21} and in space~\citep[e.g.][]{ergun19} could result in broadened current sheets due to limited spatial and/or time resolutions.

With a sizable guide field, however, ES-LHDW can be unstable inside the IDR and EDR, affecting electron and reconnection dynamics. For example, following a multi-spacecraft analysis using Cluster~\citep{norgren12}, a recent observation~\citep{chen20a} using MMS shows that strong ES-LHDW produces non-gyrotropic electron heating and vortical flows inside the EDR of reconnection with a guide field. These electron vortices have been successfully reproduced by the corresponding 3D PIC simulations~\citep{ng20} and suggest that further reconnection may occur inside the LHDW vortex tubes as dissipation at smaller scales. Other space observations of guide field reconnection show that ES-LHDW is capable of generating anomalous resistivity between electrons and ions \citep{yoo20,graham22}. 

\begin{figure}
	\centering
    \includegraphics[width=0.8\linewidth]{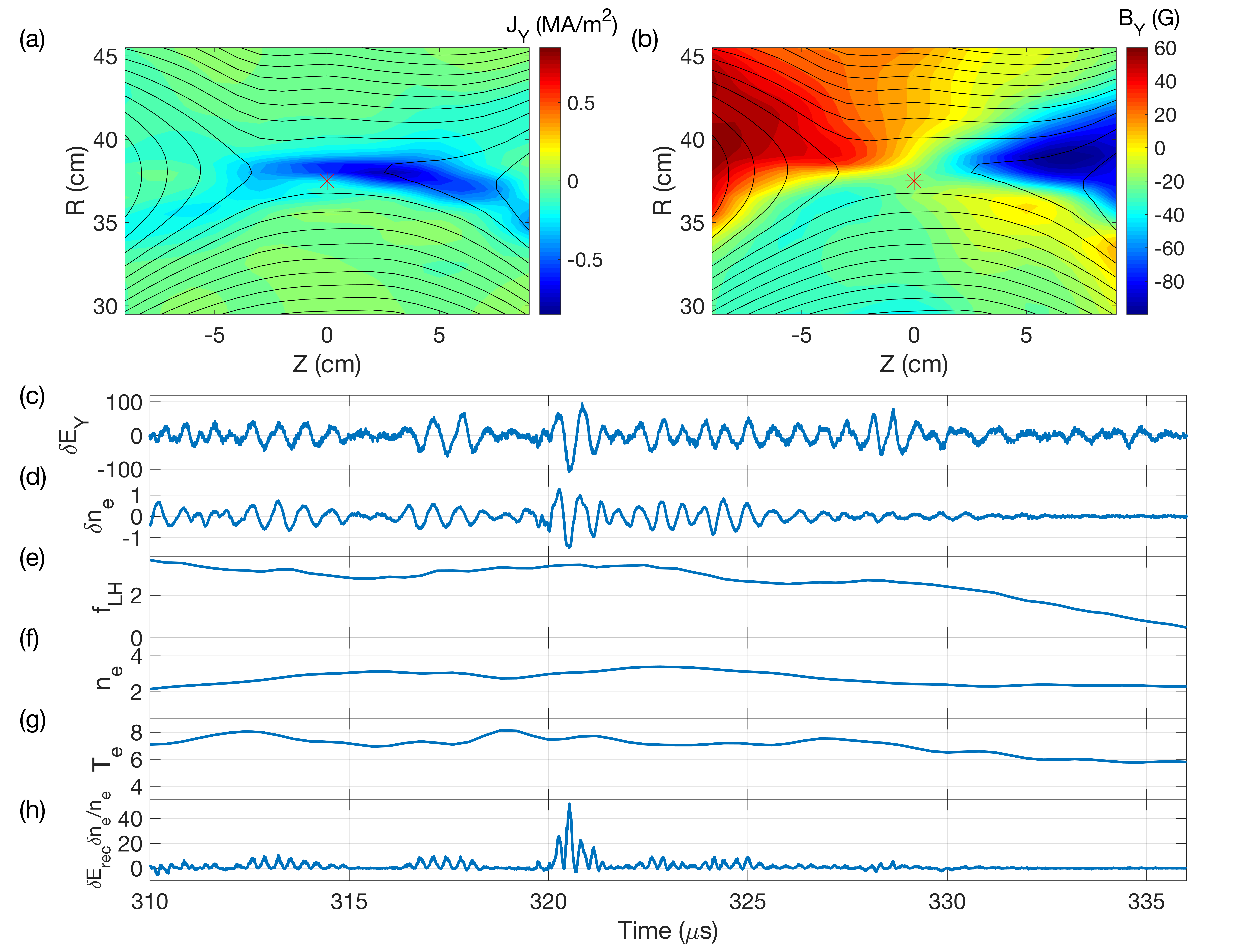}
	\caption{Measured ES-LHDW. (a,b) Out-of-plane current or magnetic field component (color) with the poloidal flux contours (black lines) representing the magnetic field lines at 326 $\mu$s. The red asterisk indicates the location of the probe. The upper side ($R>37.5$ cm) has a higher density. (c) Time series of $\delta E_{\text{rec}}$ in V/m. Wave activity near the lower hybrid frequency ($f_{\text{LH}}\sim 2$ MHz) is detected while the probe stays near the reconnection site. The amplitude of the fluctuation is comparable to the mean reconnection electric field ($\langle E_{\text{rec}} \rangle \sim 100$ V/m). (d) Time series of $\delta n_{\text{e}}$ in $10^{13}$ cm$^{-3}$ during the quasi-steady reconnection period. Time series of $f_{\text{LH}}$ (e), averaged density ($\langle n_{\text{e}} \rangle$) in $10^{13}$ cm$^{-3}$ (f), and electron temperature ($T_{\text{e}}$) in eV (g) are shown. A sharp decrease of $f_{\text{LH}}$ is observed as the approach of the X-point to the probe. Time series of $\delta E_{\text{rec}}\delta n_{\text{e}}/\langle n_{\text{e}} \rangle$ are shown in (h). Positive correlation between $\delta E_{\text{rec}}$ and $\delta n_{\text{e}}$ indicates that the wave is capable of generating anomalous resistivity. Figure from \citet{hu21}.  }
	\label{fig:mrx_lhdw}
\end{figure}

Recently, ES-LHDW measurements were revisited on MRX combined with the simultaneous measurements of electron density measurements at the same location~\citep{hu21}. Figure \ref{fig:mrx_lhdw} shows measurements of ES-LHDW at the edge of the current sheet during anti-parallel reconnection. Panels (a) and (b) show the 2D profile of the out-of-plane current density and magnetic field, respectively. The black lines are contours of the poloidal magnetic flux, representing magnetic field lines. The red asterisk is the location of the probe that measures high-frequency fluctuations in the reconnection electric field (panel c) and electron density (panel d) \citep{hu21}. Due to the positive correlation between two fluctuating quantities, the quantity of $\delta E_y\delta n_{\text{e}}/\langle n_{\text{e}} \rangle$, which is anomalous resistivity along the out-of-plane direction \citep{che11}, becomes positive. These measurements of ES-LHDW have been further extended on MRX to the cases with a sizable guide field demonstrating significant anomalous resistivity and electron heating~\citep{yoo23}. The initial corresponding 3D simulation show that ES-LHDW propagating along the outflow is triggered by the difference between electron and ion outflows in regions of low $\beta_e$~\citep{ng23}, consistent with the MRX experiment results. 

\section{Multiscale reconnection}\label{sec4}

\begin{figure}
	\centering
    \includegraphics[width=1.0\linewidth]{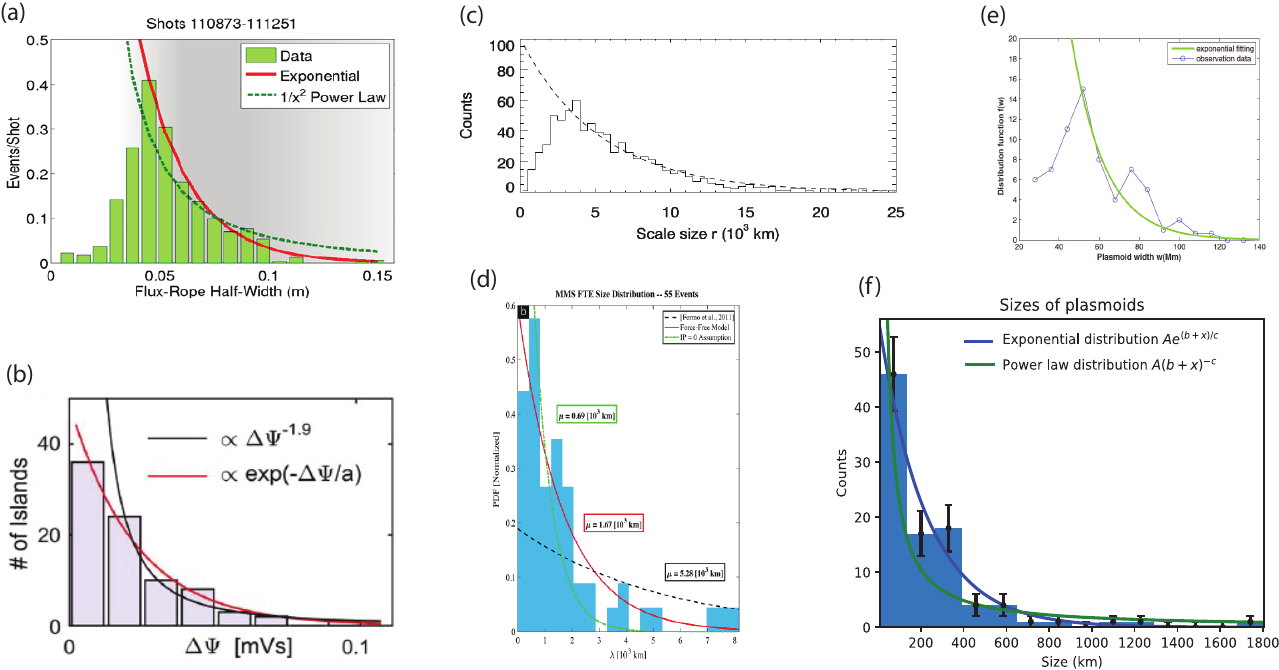}
	\caption{Plasmoid size distributions (a)~\cite{dorfman14} and (b)~\cite{olson16} from the lab; (c)~\cite{fermo11} and (d)~\cite{akhavan18} from the space observation;  (e)~\cite{guo13} from the solar observation (reproduced by permission of the AAS); and (f)~\cite{bergstedt20} from the space observation. All of them are more consistent with an exponential distribution rather than a power-law distribution.}
	\label{fig:distribution}
\end{figure}

The physic of collisionless magnetic reconnection has been studied mostly in locations nearby the local X-line as discussed in the previous sections, such as the IDR and EDR as well as separatrices. If measured in the unit of ion kinetic scales, their distances from the local X-line are not too far. However, the collisionless plasmas in space and astrophysics where reconnection is believed to occur are vastly larger - their normalized sizes have been surveyed~\citep{ji11} ranging from $\sim 10^3$ for Earth's magnetosphere to $\sim 10^{14}$ for extragalactic jets. In these large plasmas, magnetic reconnection occurs inevitably in the multiple X-line regimes as illustrated in the reconnection phase diagram~\citep{ji11,ji22preface}.

While there have been abundant evidence for collisionless multiple X-line reconnection in Earth's magnetopause as Flux Transfer Events (FTEs)~\citep{russell79} and in magnetotail as plasmoids~\citep{baker84}, there have been only relatively few laboratory work in this area~\citep{stenzel86,ono11,dorfman13,olson16,jara-almonte16}. When plasmoids form and are subsequently ejected from the current sheet, reconnection tends to proceed in an impulsive and intermittent fashion~\citep{ono11,dorfman13,jara-almonte16}, qualitatively consistent with the space observations of non-steadiness of multiscale reconnection~\citep[e.g.][]{chen08b,chen12,ergun18}.

Quantifying non-steady reconnection with multiple X-lines or ``turbulent'' reconnection is non-trivial. There have been several studies in quantifying size distributions of plasmoids, or magnetic structure in general, during multiscale reconnection, as shown in Fig.~\ref{fig:distribution}. Two are from the laboratory~\citep{dorfman14,olson16}, two from Earth's magnetopause~\citep{fermo11,akhavan18}, one from Earth's magnetotail~\citep{bergstedt20}, and one from solar observation~\citep{guo13}. Other than the last study, the others are on plasmoids on kinetic scales but all of them are more consistent with an exponential distribution rather than a power-law distribution. It is not surprising to have an exponential distribution on kinetic scales as they are dissipative scales in collisionless plasmas, but it would be a surprise if the exponential distributions also apply to fluid scales, over which the self similar power laws should apply at least in the inertial range. We note that there are interesting statistical \textit{in-situ} studies of heliospheric current sheets~\citep[e.g.][]{eriksson22} and flux ropes~\citep{janvier14} on larger scale in solar wind. The upcoming multiscale experiments, numerical simulation and observatories should shed more lights into these important questions~\citep{ji22}.

\section{Future Prospects}\label{sec5}

A concise review was given on the recent highlights from controlled laboratory studies of collisionless magnetic reconnection on a variety of topics including ion and electron kinetic structures in electromagnetic fields, energy conversion and partition, various electromagnetic and electrostatic kinetic plasma waves, as well as plasmoid-mediated multiscale reconnection. While unresolved issues still remain, many of these highlight results compare well with numerical predictions and space observations especially by the MMS mission. Thus, it is not an overstatement that the physics foundation of fast reconnection in collisionless plasmas has been largely established, at least within the parameter ranges and spatial scales that were studied.

Nonetheless, there still exist outstanding questions on the single X-line collisionless reconnection. The first question is about what dissipates magnetic fields within EDR when 2D laminar pictures do not apply. We still have cases in the laboratory where the reconnection electric field or the thickness of EDR is not fully accounted for~\citep{ji08,roytershteyn13} while in space we also have cases where 2D laminar reconnection pictures do not tell the whole story~\citep[e.g.][]{cozzani21}. Does anomalous resistivity exist in its conventional forms, as hinted by electrostatic LHDW observed during guide field reconnection~\citep{yoo23} or by IAW observed recently during anti-parallel reconnection at low ion temperature~\citep{zhang23}? Alternatively, do anomalous effects manifest as kinking of otherwise laminar 2D reconnecting current sheets~\citep{greess21} or anomalous resisitivity is cancelled by anomalous viscosity leaving no wave dissipative effects in EDR~\citep{graham22}? Further research using well-controlled experiments with adequate diagnostics, supported by matching numerical simulations, is needed to settle this long standing question.
   
Another outstanding question is about how magnetic energy is dissipated to a combination of flow, thermal and non-thermal energies of electrons and ions, as a function of field geometry, symmetry, and plasma $\beta$ at upstream. Substantial progress has been made on this subject with laboratory experiments, numerical simulations, and space observation, as summarized in Table~\ref{table:inventory} in terms of energy partition, but there remain a number of unanswered questions especially on particle acceleration. Recent progress in directly detecting accelerated electrons by reconnection electric field~\citep{chien23} and non-thermal electrons by Thomson scattering~\citep{shi22} is an encouraging sign that more results are coming. The predicted scaling of electron heating and acceleration by parallel electric field with regard to upstream $\beta$~\citep{le16} is in agreement with certain spacecraft observations~\citep{oka23}, but its laboratory study sensitively depends on plasma collisionality~\citep{le15}. High Lundquist number regimes offered by the upgraded TREX~\citep{olson16} and the upcoming Facility for Laboratory Reconnection Experiments or FLARE~\citep{ji22} will allow first laboratory accesses to the required collisionless regimes to study this important issue of collisionless reconnection.

Looking into further future, the laboratory access to multiscale regimes of magnetic reconnection is an important step as guided by the reconnection phase diagram~\citep{ji11,ji22preface}. In addition to the high Lundquist numbers, space and astrophysical plasmas have large normalized plasma system sizes, significantly expanding the parameter space over which global fluid scales and local kinetic scales are coupled. Solar corona is an excellent example where typical mean-free path of thermal particles is much longer than any kinetic scales so that locally physics is collisionless or kinetic while the mean-free path is much shorter than system sizes so that globally physics is collisional or fluid-like. How does multiscale physics across fluid and kinetic scales operate self-consistently in this regime to generate solar flares as observed, in terms of their impulsive onset and energetic consequences on the thermal heating and particle acceleration? Answering multiscale physics questions like this requires going much beyond what has been traditionally done in the reconnection research in which the detailed dynamics are studied around local X-lines based on either fluid or kinetic physics.

Statistical properties of multiscale physics need to be quantified in order to identify self-similar behavior across scales. In the case of plasmoid-mediated multiscale reconnection, despite theoretical advances in predicting power-law scaling of plasmoid sizes~\citep[e.g.][]{uzdensky10,huang12,pucci14,comisso16,majeski21}, no power-laws have been found from the laboratory or space data thus far. This may be due to the fact that data used are close to dissipative kinetic scales, and thus the accessibility of data on fluid scales are critical. To make rapid progress in this area, there exist promising opportunities to use novel techniques based on data science such as machine learning~\citep{bergstedt23} to process a huge amount of existing and new data for statistical studies. 

One of direct consequences of multiscale collisionless reconnection is its ability to accelerate particles into power-law distributions which are often observed during reconnection events. There have been a recent surge of theoretical and numerical work on this subject including reconnection under extreme conditions in astrophysics using kinetic models~\citep[e.g.][and references therein]{dahlin20,li21,guo20} and MHD models~\citep{arnold21,majeski23}, however, there has been no laboratory counterparts on this subject. It is imperative to develop new platforms~\citep[e.g.][]{chien23} for such studies as well as new diagnostics~\citep[e.g.][]{fox10,shi22} to detect accelerated non-thermal particles in the laboratory experiments, including the upcoming multiscale experiments such as FLARE~\citep{ji22}. A concerted effort from exascale modelings as well as from the scheduled or proposed multiscale space missions such as HelioSwarm~\citep{klein23} and Plasma Observatory~\citep{retino21} is critical to address these important questions.

\section*{Acknowledgement}

This work is supported in part by the NASA MMS mission. H.J. acknowledges support of this work by NASA under Grants No. NNH15AB29I and 80HQTR21T0105, and by the U.S. Department of Energy, Office of Fusion Energy Sciences under Contract No. DE-AC0209CH11466. 

\section*{Statements and Declarations}

The authors declare no competing interests.


\begin{thebibliography}{171}
\providecommand{\natexlab}[1]{#1}
\providecommand{\url}[1]{{#1}}
\providecommand{\urlprefix}{URL }
\providecommand{\doi}[1]{\url{https://doi.org/#1}}
\providecommand{\eprint}[2][]{\url{#2}}
 \bibcommenthead

\bibitem[{{Ahmadi} et~al(2022){Ahmadi}, {Eriksson}, {Newman}, {Andersson},
  {Ergun}, and {Wilder}}]{ahmadi22}
{Ahmadi} N, {Eriksson} S, {Newman} D, et~al (2022) {Observations of Electron
  Vorticity and Phase Space Holes in the Magnetopause Reconnection Separatrix}.
  Journal of Geophysical Research (Space Physics) 127(8):e30702.
  \doi{10.1029/2022JA030702}

\bibitem[{Akhavan-Tafti et~al(2018)Akhavan-Tafti, Slavin, Le, Eastwood,
  Strangeway, Russell, Nakamura, Baumjohann, Torbert, Giles, Gershman, and
  Burch}]{akhavan18}
Akhavan-Tafti M, Slavin JA, Le G, et~al (2018) Mms examination of ftes at the
  earth's subsolar magnetopause. Journal of Geophysical Research: Space Physics
  123(2):1224--1241. \doi{10.1002/2017JA024681}

\bibitem[{Arnold et~al(2021)Arnold, Drake, Swisdak, Guo, Dahlin, Chen,
  Fleishman, Glesener, Kontar, Phan, and Shen}]{arnold21}
Arnold H, Drake JF, Swisdak M, et~al (2021) Electron acceleration during
  macroscale magnetic reconnection. Phys Rev Lett 126:135,101.
  \doi{10.1103/PhysRevLett.126.135101}

\bibitem[{{Aydemir}(1992)}]{aydemir92}
{Aydemir} AY (1992) {Nonlinear studies of m=1 modes in high-temperature
  plasmas}. Physics of Fluids B 4:3469--3472. \doi{10.1063/1.860355}

\bibitem[{{Baker} et~al(1984){Baker}, {Bame}, {Birn}, {Feldman}, {Gosling},
  {Hones}, {Zwickl}, {Slavin}, {Smith}, {Tsurutani}, and {Sibeck}}]{baker84}
{Baker} DN, {Bame} SJ, {Birn} J, et~al (1984) {Direct observations of passages
  of the distant neutral line (80-140 R$_{E}$) following substorm pnsets:
  ISEE-3}. Geophys Res Lett 11(10):1042--1045. \doi{10.1029/GL011i010p01042}

\bibitem[{Bale et~al(2002)Bale, Mozer, and Phan}]{bale02}
Bale S, Mozer F, Phan T (2002) Observation of lower hybrid drift instability in
  the diffusion region at a reconnecting magnetopause. Geophys Res Lett
  29:2180. \doi{10.1029/2002GL016113}

\bibitem[{Bergerson et~al(2006)Bergerson, Forest, Fiksel, Hannum, Kendrick,
  Sarff, and Stambler}]{bergerson06}
Bergerson W, Forest C, Fiksel G, et~al (2006) Onset and saturation of the kink
  instability in a current-carrying line-tied plasma. Phys Rev Lett 96:015,004.
  \doi{10.1103/PhysRevLett.96.015004}

\bibitem[{{Bergstedt} and {Ji}(2023)}]{bergstedt23}
{Bergstedt} K, {Ji} H (2023) {A Novel Method to Train Classification Models for
  Structure Detection in In-situ Spacecraft Data}. submitted

\bibitem[{{Bergstedt} et~al(2020){Bergstedt}, {Ji}, {Jara-Almonte}, {Yoo},
  {Ergun}, and {Chen}}]{bergstedt20}
{Bergstedt} K, {Ji} H, {Jara-Almonte} J, et~al (2020) {Statistical Properties
  of Magnetic Structures and Energy Dissipation during Turbulent Reconnection
  in the Earth's Magnetotail}. Geophys Res Lett 47(19):e88540.
  \doi{10.1029/2020GL088540}

\bibitem[{{Bernstein} et~al(1957){Bernstein}, {Greene}, and
  {Kruskal}}]{bernstein57}
{Bernstein} IB, {Greene} JM, {Kruskal} MD (1957) {Exact Nonlinear Plasma
  Oscillations}. Physical Review 108(3):546--550. \doi{10.1103/PhysRev.108.546}

\bibitem[{Birn et~al(2001)Birn, Drake, Shay, Rogers, Denton, Hesse, Kuznetsova,
  Ma, Bhattachargee, Otto, and Pritchett}]{birn01}
Birn J, Drake J, Shay M, et~al (2001) Geomagnetic {E}nvironmental {M}odeling
  ({GEM}) {M}agnetic {R}econnection {C}hallenge. J Geophys Res 106(A3):3715.
  \doi{10.1029/1999JA900449}

\bibitem[{Biskamp et~al(1995)Biskamp, Schwarz, and Drake}]{biskamp95}
Biskamp D, Schwarz E, Drake J (1995) Ion-controlled collisionless magnetic
  reconnection. Phys Rev Lett 75:3850. \doi{10.1103/PhysRevLett.75.3850}

\bibitem[{Bose et~al(2022)Bose, Fox, Ji, Yoo, Goodman, Alt, Jara-Almonte, and
  Yamada}]{bose22}
Bose S, Fox W, Ji H, et~al (2022) Conversion of magnetic energy to plasma
  kinetic energy during guide field magnetic reconnection in the laboratory.
  submitted

\bibitem[{Bowers et~al(2009)Bowers, Albright, Yin, Daughton, Roytershteyn,
  Bergen, and Kwan}]{bowers09}
Bowers K, Albright B, Yin L, et~al (2009) Advances in petascale kinetic
  simulations with {VPIC} and {R}oadrunner. Journal of Physics: Conference
  Series 180:012,055. \doi{10.1088/1742-6596/180/1/012055}

\bibitem[{{Bratenahl} and {Yeates}(1970)}]{bratenahl70}
{Bratenahl} A, {Yeates} CM (1970) {Experimental Study of Magnetic Flux Transfer
  at the Hyperbolic Neutral Point}. Phys Fluids 13:2696--2709.
  \doi{10.1063/1.1692853}

\bibitem[{Brown(1999)}]{brown99}
Brown M (1999) Experimental studies of magnetic reconnection. Phys Plasmas
  6:1717. \doi{10.1063/1.873430}

\bibitem[{Brown et~al(2002)Brown, Cothran, Landreman, Schlossberg, and
  Matthaeus}]{brown02}
Brown M, Cothran C, Landreman M, et~al (2002) Experimental observation of
  energetic ions accelerated by three-dimensional magnetic reconnection in a
  laboratory plasma. Astrophys J 577:L63. \doi{10.1086/344145}

\bibitem[{Brown et~al(2006)Brown, Cothran, and Fung}]{brown06}
Brown MR, Cothran CD, Fung J (2006) Two fluid effects on three-dimensional
  reconnection in the swarthmore spheromak experiment with comparisons to space
  data. Phys Plasmas 13(5):056,503. \doi{10.1063/1.2180729}

\bibitem[{{Burch} et~al(2016){Burch}, {Torbert}, {Phan}, {Chen}, {Moore},
  {Ergun}, {Eastwood}, {Gershman}, {Cassak}, {Argall}, {Wang}, {Hesse},
  {Pollock}, {Giles}, {Nakamura}, {Mauk}, {Fuselier}, {Russell}, {Strangeway},
  {Drake}, {Shay}, {Khotyaintsev}, {Lindqvist}, {Marklund}, {Wilder}, {Young},
  {Torkar}, {Goldstein}, {Dorelli}, {Avanov}, {Oka}, {Baker}, {Jaynes},
  {Goodrich}, {Cohen}, {Turner}, {Fennell}, {Blake}, {Clemmons}, {Goldman},
  {Newman}, {Petrinec}, {Trattner}, {Lavraud}, {Reiff}, {Baumjohann}, {Magnes},
  {Steller}, {Lewis}, {Saito}, {Coffey}, and {Chandler}}]{burch16}
{Burch} JL, {Torbert} RB, {Phan} TD, et~al (2016) {Electron-scale measurements
  of magnetic reconnection in space}. Science 352:aaf2939.
  \doi{10.1126/science.aaf2939}

\bibitem[{{Cai} and {Lee}(1997)}]{cai97}
{Cai} HJ, {Lee} LC (1997) {The generalized Ohm's law in collisionless magnetic
  reconnection}. Phys Plasmas 4:509. \doi{10.1063/1.872178}

\bibitem[{Carter et~al(2001)Carter, Ji, Trintchouk, Yamada, and
  Kulsrud}]{carter01}
Carter T, Ji H, Trintchouk F, et~al (2001) Measurement of lower-hybrid drift
  turbulence in a reconnecting current sheet. Phys Rev Lett 88:015,001.
  \doi{10.1103/PhysRevLett.88.015001}

\bibitem[{Cattell et~al(2005)Cattell, Dombeck, Wygant, Drake, Swisdak,
  Goldstein, Keith, Fazakerley, Andre, Lucek, and Balogh}]{cattell05}
Cattell C, Dombeck J, Wygant J, et~al (2005) Cluster observations of electron
  holes in association with magnetotail reconnection and comparison to
  simulations. J Geophys Res 110:A01,211. \doi{10.1029/2004JA010519}

\bibitem[{Cerutti et~al(2013)Cerutti, Werner, Uzdensky, and
  Begelman}]{cerutti13}
Cerutti B, Werner GR, Uzdensky DA, et~al (2013) {Simulations of Particle
  Acceleration beyond the Classical Synchrotron Burnoff Limit in Magnetic
  Reconnection: An Explanation of the Crab Flares}. Astrophys J 770(2):147.
  \doi{10.1088/0004-637X/770/2/147}

\bibitem[{Che et~al(2011)Che, Drake, and Swisdak}]{che11}
Che H, Drake JF, Swisdak M (2011) A current filamentation mechanism for
  breaking magnetic field lines during reconnection. Nature 474(7350):184--187.
  \doi{10.1038/nature10091}

\bibitem[{Chen et~al(2008)Chen, Bessho, Lefebvre, Vaith, Fazakerley,
  Bhattacharjee, Puhl-Quinn, Runov, Khotyaintsev, Vaivads, Georgescu, and
  Torbert}]{chen08b}
Chen LJ, Bessho N, Lefebvre B, et~al (2008) Evidence of an extended electron
  current sheet and its neighboring magnetic island during magnetotail
  reconnection. J Geophys Res 113:A12,213. \doi{10.1029/2008JA013385}

\bibitem[{Chen et~al(2012)Chen, Daughton, Bhattacharjee, Torbert, Roytershteyn,
  and Bessho}]{chen12}
Chen LJ, Daughton W, Bhattacharjee A, et~al (2012) In-plane electric fields in
  magnetic islands during collisionless magnetic reconnection. Physics of
  Plasmas 19(11):112,902. \doi{10.1063/1.4767645}

\bibitem[{{Chen} et~al(2020){Chen}, {Wang}, {Le Contel}, {Rager}, {Hesse},
  {Drake}, {Dorelli}, {Ng}, {Bessho}, {Graham}, {Wilson}, {Moore}, {Giles},
  {Paterson}, {Lavraud}, {Genestreti}, {Nakamura}, {Khotyaintsev}, {Ergun},
  {Torbert}, {Burch}, {Pollock}, {Russell}, {Lindqvist}, and
  {Avanov}}]{chen20a}
{Chen} LJ, {Wang} S, {Le Contel} O, et~al (2020) {Lower-Hybrid Drift Waves
  Driving Electron Nongyrotropic Heating and Vortical Flows in a Magnetic
  Reconnection Layer}. Phys Rev Lett 125(2):025103.
  \doi{10.1103/PhysRevLett.125.025103}

\bibitem[{Chien et~al(2023)Chien, Gao, Zhang, Ji, Blackman, Daughton, Stanier,
  Le, Guo, Follett, Chen, Fiksel, Bleotu, Cauble, Chen, Fazzini, Flippo,
  French, Froula, Fuchs, Fujioka, Hill, Klein, Kuranz, Nilson, Rasmus, and
  Takizawa}]{chien23}
Chien A, Gao L, Zhang S, et~al (2023) Non-thermal electron acceleration from
  magnetically driven reconnection in a laboratory plasma. Nature Physics 19.
  \doi{10.1038/s41567-022-01839-x}

\bibitem[{{Comisso} et~al(2016){Comisso}, {Lingam}, {Huang}, and
  {Bhattacharjee}}]{comisso16}
{Comisso} L, {Lingam} M, {Huang} YM, et~al (2016) {General theory of the
  plasmoid instability}. Phys Plasmas 23(10):100702. \doi{10.1063/1.4964481}

\bibitem[{{Cozzani} et~al(2021){Cozzani}, {Khotyaintsev}, {Graham}, {Egedal},
  {Andr{\'e}}, {Vaivads}, {Alexandrova}, {Le Contel}, {Nakamura}, {Fuselier},
  {Russell}, and {Burch}}]{cozzani21}
{Cozzani} G, {Khotyaintsev} YV, {Graham} DB, et~al (2021) {Structure of a
  Perturbed Magnetic Reconnection Electron Diffusion Region in the Earth's
  Magnetotail}. Phys Rev Lett 127(21):215101.
  \doi{10.1103/PhysRevLett.127.215101}

\bibitem[{Dahlin(2020)}]{dahlin20}
Dahlin JT (2020) Prospectus on electron acceleration via magnetic reconnection.
  Phys Plasmas 27(10):100,601. \doi{10.1063/5.0019338}

\bibitem[{Daughton(2003)}]{daughton03}
Daughton W (2003) Electromagnetic properties of the lower-hybrid drift
  instability in a thin current sheet. Phys Plasmas 10:3103.
  \doi{10.1063/1.1594724}

\bibitem[{{Daughton} et~al(2004){Daughton}, {Lapenta}, and
  {Ricci}}]{daughton04}
{Daughton} W, {Lapenta} G, {Ricci} P (2004) {Nonlinear Evolution of the
  Lower-Hybrid Drift Instability in a Current Sheet}. Phys Rev Lett
  93(10):105,004. \doi{10.1103/PhysRevLett.93.105004}

\bibitem[{Davidson and Gladd(1975)}]{davidson75}
Davidson R, Gladd N (1975) Anomalous transport properties associated with the
  lower-hybrid drift instability. Phys Fluids 18:1327. \doi{10.1063/1.861021}

\bibitem[{Dorfman et~al(2013)Dorfman, Ji, Yamada, Yoo, Lawrence, Myers, and
  Tharp}]{dorfman13}
Dorfman S, Ji H, Yamada M, et~al (2013) Three-dimensional, impulsive magnetic
  reconnection in a laboratory plasma. Geophys Res Lett 40:233--238.
  \doi{10.1029/2012GL054574}

\bibitem[{{Dorfman} et~al(2014){Dorfman}, {Ji}, {Yamada}, {Yoo}, {Lawrence},
  {Myers}, and {Tharp}}]{dorfman14}
{Dorfman} S, {Ji} H, {Yamada} M, et~al (2014) {Experimental observation of 3-D,
  impulsive reconnection events in a laboratory plasma}. Physics of Plasmas
  21(1):012109. \doi{10.1063/1.4862039}

\bibitem[{{Drake} et~al(2006){Drake}, {Swisdak}, {Che}, and {Shay}}]{drake06}
{Drake} JF, {Swisdak} M, {Che} H, et~al (2006) {Electron acceleration from
  contracting magnetic islands during reconnection}. Nature 443:553--556.
  \doi{10.1038/nature05116}

\bibitem[{Dungey(1961)}]{dungey61}
Dungey J (1961) Interplanetary magnetic field and the auroral zones. Phys Rev
  Lett 6(2):47. \doi{10.1103/PhysRevLett.6.47}

\bibitem[{Eastwood et~al(2010)Eastwood, Phan, Øieroset, and
  Shay}]{eastwood10b}
Eastwood JP, Phan TD, Øieroset M, et~al (2010) Average properties of the
  magnetic reconnection ion diffusion region in the earth's magnetotail: The
  2001-2005 cluster observations and comparison with simulations. Journal of
  Geophysical Research: Space Physics 115. \doi{10.1029/2009JA014962}

\bibitem[{{Eastwood} et~al(2013){Eastwood}, {Phan}, {Drake}, {Shay}, {Borg},
  {Lavraud}, and {Taylor}}]{eastwood13}
{Eastwood} JP, {Phan} TD, {Drake} JF, et~al (2013) {Energy Partition in
  Magnetic Reconnection in Earth's Magnetotail}. Phys Rev Lett 110(22):225001.
  \doi{10.1103/PhysRevLett.110.225001}

\bibitem[{{Eastwood} et~al(2020){Eastwood}, {Goldman}, {Phan}, {Stawarz},
  {Cassak}, {Drake}, {Newman}, {Lavraud}, {Shay}, {Ergun}, {Burch}, {Gershman},
  {Giles}, {Lindqvist}, {Torbert}, {Strangeway}, and {Russell}}]{eastwood20}
{Eastwood} JP, {Goldman} MV, {Phan} TD, et~al (2020) {Energy Flux Densities
  near the Electron Dissipation Region in Asymmetric Magnetopause
  Reconnection}. Phys Rev Lett 125(26):265102.
  \doi{10.1103/PhysRevLett.125.265102}

\bibitem[{Egedal and Fasoli(2001)}]{egedal01}
Egedal J, Fasoli A (2001) Single-particle dynamics in collisionless magnetic
  reconnection. Phys Rev Lett 86(22):5047. \doi{10.1103/PhysRevLett.86.5047}

\bibitem[{{Egedal} et~al(2000){Egedal}, {Fasoli}, {Porkolab}, and
  {Tarkowski}}]{egedal00}
{Egedal} J, {Fasoli} A, {Porkolab} M, et~al (2000) {Plasma generation and
  confinement in a toroidal magnetic cusp}. Rev Sci Instr 71:3351--3361.
  \doi{10.1063/1.1287340}

\bibitem[{Egedal et~al(2003)Egedal, Fasoli, and Nazemi}]{egedal03}
Egedal J, Fasoli A, Nazemi J (2003) Dynamical plasma response during driven
  magnetic reconnection. Phys Rev Lett 90:135,003.
  \doi{10.1103/PhysRevLett.90.135003}

\bibitem[{Egedal et~al(2013)Egedal, Le, and Daughton}]{egedal13}
Egedal J, Le A, Daughton W (2013) A review of pressure anisotropy caused by
  electron trapping in collisionless plasma, and its implications for magnetic
  reconnection. Physics of Plasmas 20(6):061,201. \doi{10.1063/1.4811092}

\bibitem[{Egedal et~al(2018)Egedal, Le, Daughton, Wetherton, Cassak, Burch,
  Lavraud, Dorelli, Gershman, and Avanov}]{egedal18}
Egedal J, Le A, Daughton W, et~al (2018) Spacecraft observations of oblique
  electron beams breaking the frozen-in law during asymmetric reconnection.
  Physical Review Letters 120:055,101. \doi{10.1103/PhysRevLett.120.055101}

\bibitem[{Egedal et~al(2019)Egedal, Ng, Le, Daughton, Wetherton, Dorelli,
  Gershman, and Rager}]{egedal19}
Egedal J, Ng J, Le A, et~al (2019) Pressure tensor elements breaking the
  frozen-in law during reconnection in earth's magnetotail. Physical Review
  Letters 123. \doi{10.1103/PhysRevLett.123.225101}

\bibitem[{Ergun et~al(2016)Ergun, Goodrich, Wilder, Holmes, Stawarz, Eriksson,
  Sturner, Malaspina, Usanova, Torbert, Lindqvist, Khotyaintsev, Burch,
  Strangeway, Russell, Pollock, Giles, Hesse, Chen, Lapenta, Goldman, Newman,
  Schwartz, Eastwood, Phan, Mozer, Drake, Shay, Cassak, Nakamura, and
  Marklund}]{ergun16a}
Ergun RE, Goodrich KA, Wilder FD, et~al (2016) {Magnetospheric Multiscale
  Satellites Observations of Parallel Electric Fields Associated with Magnetic
  Reconnection}. Phys Rev Lett 116(2):235,102.
  \doi{10.1103/PhysRevLett.116.235102}

\bibitem[{{Ergun} et~al(2016){Ergun}, {Holmes}, {Goodrich}, {Wilder},
  {Stawarz}, {Eriksson}, {Newman}, {Schwartz}, {Goldman}, {Sturner},
  {Malaspina}, {Usanova}, {Torbert}, {Argall}, {Lindqvist}, {Khotyaintsev},
  {Burch}, {Strangeway}, {Russell}, {Pollock}, {Giles}, {Dorelli}, {Avanov},
  {Hesse}, {Chen}, {Lavraud}, {Le Contel}, {Retino}, {Phan}, {Eastwood},
  {Oieroset}, {Drake}, {Shay}, {Cassak}, {Nakamura}, {Zhou}, {Ashour-Abdalla},
  and {Andr{\'e}}}]{ergun16b}
{Ergun} RE, {Holmes} JC, {Goodrich} KA, et~al (2016) {Magnetospheric Multiscale
  observations of large-amplitude, parallel, electrostatic waves associated
  with magnetic reconnection at the magnetopause}. Geophys Res Lett
  43:5626--5634. \doi{10.1002/2016GL068992}

\bibitem[{Ergun et~al(2017)Ergun, Chen, Wilder, Ahmadi, Eriksson, Usanova,
  Goodrich, Holmes, Sturner, Malaspina, Newman, Torbert, Argall, Lindqvist,
  Burch, Webster, Drake, Price, Cassak, Swisdak, Shay, Graham, Strangeway,
  Russell, Giles, Dorelli, Gershman, Avanov, Hesse, Lavraud, Le~Contel,
  Retin{\`o}, Phan, Goldman, Stawarz, Schwartz, Eastwood, Hwang, Nakamura, and
  Wang}]{ergun17}
Ergun RE, Chen LJ, Wilder FD, et~al (2017) {Drift waves, intense parallel
  electric fields, and turbulence associated with asymmetric magnetic
  reconnection at the magnetopause}. Geophys Res Lett 44(7):2978--2986.
  \doi{10.1002/2016GL072493}

\bibitem[{{Ergun} et~al(2018){Ergun}, {Goodrich}, {Wilder}, {Ahmadi}, {Holmes},
  {Eriksson}, {Stawarz}, {Nakamura}, {Genestreti}, {Hesse}, {Burch}, {Torbert},
  {Phan}, {Schwartz}, {Eastwood}, {Strangeway}, {Le Contel}, {Russell},
  {Argall}, {Lindqvist}, {Chen}, {Cassak}, {Giles}, {Dorelli}, {Gershman},
  {Leonard}, {Lavraud}, {Retino}, {Matthaeus}, and {Vaivads}}]{ergun18}
{Ergun} RE, {Goodrich} KA, {Wilder} FD, et~al (2018) {Magnetic Reconnection,
  Turbulence, and Particle Acceleration: Observations in the Earth's
  Magnetotail}. Geophys Res Lett 45(8):3338--3347. \doi{10.1002/2018GL076993}

\bibitem[{Ergun et~al(2019)Ergun, Hoilijoki, Ahmadi, Schwartz, Wilder, Drake,
  Hesse, Shay, Ji, Yamada, Graham, Cassak, Swisdak, Burch, Torbert, Holmes,
  Stawarz, Goodrich, Eriksson, Strangeway, and LeContel}]{ergun19}
Ergun RE, Hoilijoki S, Ahmadi N, et~al (2019) Magnetic reconnection in three
  dimensions: Modeling and analysis of electromagnetic drift waves in the
  adjacent current sheet. Journal of Geophysical Research: Space Physics
  124:10,085--10,103. \doi{10.1029/2019JA027275}

\bibitem[{Eriksson et~al(2016)Eriksson, Wilder, Ergun, Schwartz, Cassak, Burch,
  Chen, Torbert, Phan, Lavraud, Goodrich, Holmes, Stawarz, Sturner, Malaspina,
  Usanova, Trattner, Strangeway, Russell, Pollock, Giles, Hesse, Lindqvist,
  Drake, Shay, Nakamura, and Marklund}]{eriksson16}
Eriksson S, Wilder FD, Ergun RE, et~al (2016) {Magnetospheric Multiscale
  Observations of the Electron Diffusion Region of Large Guide Field Magnetic
  Reconnection}. Phys Rev Lett 117(1):015,001.
  \doi{10.1103/PhysRevLett.117.015001}

\bibitem[{Eriksson et~al(2022)Eriksson, Swisdak, Weygand, Mallet, Newman,
  Lapenta, III, Turner, and Larsen}]{eriksson22}
Eriksson S, Swisdak M, Weygand JM, et~al (2022) Characteristics of multi-scale
  current sheets in the solar wind at 1 au associated with magnetic
  reconnection and the case for a heliospheric current sheet avalanche. The
  Astrophysical Journal 933:181. \doi{10.3847/1538-4357/ac73f6}

\bibitem[{{Fermo} et~al(2011){Fermo}, {Drake}, {Swisdak}, and
  {Hwang}}]{fermo11}
{Fermo} RL, {Drake} JF, {Swisdak} M, et~al (2011) {Comparison of a statistical
  model for magnetic islands in large current layers with Hall MHD simulations
  and Cluster FTE observations}. J Geophys Res 116:A09226.
  \doi{10.1029/2010JA016271}

\bibitem[{{Fox} et~al(2008){Fox}, {Porkolab}, {Egedal}, {Katz}, and Le}]{fox08}
{Fox} W, {Porkolab} M, {Egedal} J, et~al (2008) {Laboratory Observation of
  Electron Phase-Space Holes During Magnetic Reconnection}. Phys Rev Lett
  101:255,003. \doi{10.1103/PhysRevLett.101.255003}

\bibitem[{Fox et~al(2010)Fox, Porkolab, Egedal, Katz, , and Le}]{fox10}
Fox W, Porkolab M, Egedal J, et~al (2010) {Laboratory observations of electron
  energization and associated lower-hybrid and Trivelpiece-Gould wave
  turbulence during magnetic reconnection}. Phys Plasmas 17:072,303.
  \doi{10.1063/1.3435216}

\bibitem[{Fox et~al(2012)Fox, Porkolab, Egedal, Katz, and Le}]{fox12a}
Fox W, Porkolab M, Egedal J, et~al (2012) Observations of electron phase-space
  holes driven during magnetic reconnection in a laboratory plasma. Phys
  Plasmas 19:032,118. \doi{10.1063/1.3692224}

\bibitem[{Fox et~al(2017)Fox, Sciortino, von Stechow, Jara-Almonte, Yoo, Ji,
  and Yamada}]{fox17}
Fox W, Sciortino F, von Stechow A, et~al (2017) Experimental verification of
  the role of electron pressure in fast magnetic reconnection with a guide
  field. Phys Rev Lett 118:125,002. \doi{10.1103/PhysRevLett.118.125002}

\bibitem[{{Fox} et~al(2018){Fox}, {Wilder}, {Eriksson}, {Jara-Almonte},
  {Pucci}, {Yoo}, {Ji}, {Yamada}, {Ergun}, {Oieroset}, and {Phan}}]{fox18}
{Fox} W, {Wilder} FD, {Eriksson} S, et~al (2018) {Energy Conversion by Parallel
  Electric Fields During Guide Field Reconnection in Scaled Laboratory and
  Space Experiments}. Geophys Res Lett 45(23):12,677--12,684.
  \doi{10.1029/2018GL079883}

\bibitem[{{Furno} et~al(2007){Furno}, {Intrator}, {Lapenta}, {Dorf}, {Abbate},
  and {Ryutov}}]{furno07}
{Furno} I, {Intrator} TP, {Lapenta} G, et~al (2007) {Effects of boundary
  conditions and flow on the kink instability in a cylindrical plasma column}.
  Phys Plasmas 14:022,103. \doi{10.1063/1.2435306}

\bibitem[{Gekelman and Stenzel(1984)}]{gekelman84}
Gekelman W, Stenzel R (1984) Magnetic field line reconnection experiments: 6.
  magnetic turbulence. Journal of Geophysical Research: Space Physics
  89(A5):2715--2733. \doi{10.1029/JA089iA05p02715}

\bibitem[{Gekelman and Stenzel(1985)}]{gekelman85}
Gekelman W, Stenzel R (1985) Measurement and instability analysis of
  three-dimensional anisotropic electron distribution functions. Phys Rev Lett
  54(22):2414. \doi{10.1103/PhysRevLett.54.2414}

\bibitem[{Graham et~al(2023)Graham, Khotyaintsev, Cozzani, Wilder, Homes,
  Nakamura, Buechner, Dokgo, Richard, Steinvall, Norgren, Chen, Drake, Ji,
  Stawarz, and Eriksson}]{graham23}
Graham D, Khotyaintsev Y, Cozzani G, et~al (2023) The role of kinetic
  instabilities and waves in collisionless magnetic reconnection. in
  preparation

\bibitem[{{Graham} et~al(2022){Graham}, {Khotyaintsev}, {Andr{\'e}}, {Vaivads},
  {Divin}, {Drake}, {Norgren}, {Le Contel}, {Lindqvist}, {Rager}, {Gershman},
  {Russell}, {Burch}, {Hwang}, and {Dokgo}}]{graham22}
{Graham} DB, {Khotyaintsev} YV, {Andr{\'e}} M, et~al (2022) {Direct
  observations of anomalous resistivity and diffusion in collisionless plasma}.
  Nature Communications 13:2954. \doi{10.1038/s41467-022-30561-8}

\bibitem[{{Greess} et~al(2021){Greess}, {Egedal}, {Stanier}, {Daughton},
  {Olson}, {L{\^e}}, {Myers}, {Millet-Ayala}, {Clark}, {Wallace}, {Endrizzi},
  and {Forest}}]{greess21}
{Greess} S, {Egedal} J, {Stanier} A, et~al (2021) {Laboratory Verification of
  Electron-Scale Reconnection Regions Modulated by a Three-Dimensional
  Instability}. Journal of Geophysical Research (Space Physics) 126(7):e29316.
  \doi{10.1029/2021JA029316}

\bibitem[{Guo et~al(2020)Guo, Liu, Li, Li, Daughton, and Kilian}]{guo20}
Guo F, Liu YH, Li X, et~al (2020) Recent progress on particle acceleration and
  reconnection physics during magnetic reconnection in the
  magnetically-dominated relativistic regime. Phys Plasmas 27(8):080,501.
  \doi{10.1063/5.0012094}

\bibitem[{{Guo} et~al(2013){Guo}, {Bhattacharjee}, and {Huang}}]{guo13}
{Guo} LJ, {Bhattacharjee} A, {Huang} YM (2013) {Distribution of Plasmoids in
  Post-coronal Mass Ejection Current Sheets}. Astrophys J Lett 771:L14.
  \doi{10.1088/2041-8205/771/1/L14}

\bibitem[{Hesse et~al(1999)Hesse, Schindler, Birn, and Kuznetsova}]{hesse99}
Hesse M, Schindler K, Birn J, et~al (1999) The diffusion region in
  collisionless magnetic reconnection. Phys Plasmas 6:1781.
  \doi{10.1063/1.873436}

\bibitem[{Hesse et~al(2014)Hesse, Aunai, Sibeck, and Birn}]{hesse14}
Hesse M, Aunai N, Sibeck D, et~al (2014) On the electron diffusion region in
  planar, asymmetric, systems. Geophysical Research Letters 41(24):8673--8680.
  \doi{10.1002/2014GL061586}

\bibitem[{{Hoshino} et~al(2001){Hoshino}, {Mukai}, {Terasawa}, and
  {Shinohara}}]{hoshino01}
{Hoshino} M, {Mukai} T, {Terasawa} T, et~al (2001) {Suprathermal electron
  acceleration in magnetic reconnection}. J Geophys Res
  106(A11):25,979--25,998. \doi{10.1029/2001JA900052}

\bibitem[{Hsu et~al(2000)Hsu, Fiksel, Carter, Ji, Kulsrud, and Yamada}]{hsu00}
Hsu S, Fiksel G, Carter T, et~al (2000) Local measurement of nonclassical ion
  heating during magnetic reconnection. Phys Rev Lett 84:3859.
  \doi{10.1103/PhysRevLett.84.3859}

\bibitem[{Hu et~al(2021)Hu, Yoo, Ji, Goodman, and Wu}]{hu21}
Hu Y, Yoo J, Ji H, et~al (2021) Probe measurements of electric field and
  electron density fluctuations at megahertz frequencies using in-shaft
  miniature circuits. Review of Scientific Instruments 92.
  \doi{10.1063/5.0035135}

\bibitem[{Huang and Bhattacharjee(2012)}]{huang12}
Huang YM, Bhattacharjee A (2012) {Distribution of Plasmoids in
  High-Lundquist-Number Magnetic Reconnection}. Phys Rev Lett 109(26):265,002.
  \doi{10.1103/PhysRevLett.109.265002}

\bibitem[{{Janvier} et~al(2014){Janvier}, {D{\'e}moulin}, and
  {Dasso}}]{janvier14}
{Janvier} M, {D{\'e}moulin} P, {Dasso} S (2014) {In situ properties of small
  and large flux ropes in the solar wind}. Journal of Geophysical Research
  (Space Physics) 119(9):7088--7107. \doi{10.1002/2014JA020218}

\bibitem[{{Jara-Almonte} et~al(2014){Jara-Almonte}, {Daughton}, and
  {Ji}}]{jara-almonte14}
{Jara-Almonte} J, {Daughton} W, {Ji} H (2014) {Debye scale turbulence within
  the electron diffusion layer during magnetic reconnection}. Phys Plasmas
  21(3):032114. \doi{10.1063/1.4867868}

\bibitem[{Jara-Almonte et~al(2016)Jara-Almonte, Ji, Yamada, Yoo, and
  Fox}]{jara-almonte16}
Jara-Almonte J, Ji H, Yamada M, et~al (2016) {Laboratory Observation of
  Resistive Electron Tearing in a Two-Fluid Reconnecting Current Sheet}.
  Physical Review Letters 117(9):095,001. \doi{10.1103/PhysRevLett.117.095001}

\bibitem[{Ji and Daughton(2011)}]{ji11}
Ji H, Daughton W (2011) Phase diagram for magnetic reconnection in
  heliophysical, astrophysical, and laboratory plasmas. Phys Plasmas
  18(11):111207. \doi{10.1063/1.3647505}

\bibitem[{Ji and Daughton(2022)}]{ji22preface}
Ji H, Daughton W (2022) Preface for frontiers of magnetic reconnection research
  in heliophysical, astrophysical, and laboratory plasmas. Physics of Plasmas
  29(7):070,401. \doi{10.1063/5.0104925}

\bibitem[{Ji et~al(1998)Ji, Yamada, Hsu, and Kulsrud}]{ji98}
Ji H, Yamada M, Hsu S, et~al (1998) Experimental test of the sweet-parker model
  of magnetic reconnection. Phys Rev Lett 80:3256.
  \doi{10.1103/PhysRevLett.80.3256}

\bibitem[{Ji et~al(2004)Ji, Terry, Yamada, Kulsrud, Kuritsyn, and Ren}]{ji04}
Ji H, Terry S, Yamada M, et~al (2004) Electromagnetic fluctuation during fast
  reconnection in a laboratory plasma. Phys Rev Lett 92:115,001.
  \doi{10.1103/PhysRevLett.92.115001}

\bibitem[{Ji et~al(2005)Ji, Kulsrud, Fox, and Yamada}]{ji05}
Ji H, Kulsrud R, Fox W, et~al (2005) An obliquely propagating electromagnetic
  drift instability in the lower hybrid frequency range. J Geophys Res
  110:A08,212. \doi{10.1029/2005JA011188}

\bibitem[{{Ji} et~al(2008){Ji}, {Ren}, {Yamada}, {Dorfman}, {Daughton}, and
  {Gerhardt}}]{ji08}
{Ji} H, {Ren} Y, {Yamada} M, et~al (2008) {New insights into dissipation in the
  electron layer during magnetic reconnection}. Geophys Res Lett 35:L13,106.
  \doi{10.1029/2008GL034538}

\bibitem[{Ji et~al(2022)Ji, Daughton, Jara-Almonte, Le, Stanier, and
  Yoo}]{ji22}
Ji H, Daughton W, Jara-Almonte J, et~al (2022) Magnetic reconnection in the era
  of exascale computing and multiscale experiments. Nat Rev Phys 4:263--282.
  \doi{10.1038/s42254-021-00419-x}

\bibitem[{Katz et~al(2010)Katz, Egedal, Fox, Le, Bonde, and
  Vrublevskis}]{katz10}
Katz N, Egedal J, Fox W, et~al (2010) Laboratory observation of localized onset
  of magnetic reconnection. Phys Rev Lett 104(25):255,004.
  \doi{10.1103/PhysRevLett.104.255004}

\bibitem[{Khotyaintsev et~al(2019)Khotyaintsev, Graham, Norgren, and
  Vaivads}]{khotyaintsev19}
Khotyaintsev YV, Graham DB, Norgren C, et~al (2019) Collisionless magnetic
  reconnection and waves: Progress review. Frontiers in Astronomy and Space
  Sciences 6:70. \doi{10.3389/fspas.2019.00070}

\bibitem[{Khotyaintsev et~al(2020)Khotyaintsev, Graham, Steinvall, Alm,
  Vaivads, Johlander, Norgren, Li, Divin, Fu, Hwang, Burch, Ahmadi, Contel,
  Gershman, Russell, and Torbert}]{khotyaintsev20}
Khotyaintsev YV, Graham DB, Steinvall K, et~al (2020) Electron heating by
  debye-scale turbulence in guide-field reconnection. Physical Review Letters
  124. \doi{10.1103/PhysRevLett.124.045101}

\bibitem[{{Klein} et~al(2023){Klein}, {Spence}, {Alexandrova}, {Argall},
  {Arzamasskiy}, {Bookbinder}, {Broeren}, {Caprioli}, {Case}, {Chandran},
  {Chen}, {Dors}, {Eastwood}, {Forsyth}, {Galvin}, {Genot}, {Halekas}, {Hesse},
  {Hine}, {Horbury}, {Jian}, {Kasper}, {Kretzschmar}, {Kunz}, {Lavraud}, {Le
  Contel}, {Mallet}, {Maruca}, {Matthaeus}, {Niehof}, {O'Brian}, {Owen},
  {Retino}, {Reynolds}, {Roberts}, {Schekochihin}, {Skoug}, {Smith}, {Smith},
  {Steinberg}, {Stevens}, {Szabo}, {TenBarge}, {Torbert}, {Vasquez},
  {Verscharen}, {Whittlesey}, {Wickizer}, {Zank}, and {Zweibel}}]{klein23}
{Klein} KG, {Spence} H, {Alexandrova} O, et~al (2023) {HelioSwarm: A
  Multipoint, Multiscale Mission to Characterize Turbulence}. arXiv e-prints
  arXiv:2306.06537. \doi{10.48550/arXiv.2306.06537}

\bibitem[{Kleva et~al(1995)Kleva, Drake, and Waelbroeck}]{kleva95}
Kleva R, Drake J, Waelbroeck F (1995) Fast reconnection in high temperature
  plasmas. Phys Plasmas 2:23. \doi{10.1063/1.871095}

\bibitem[{Krall and Liewer(1971)}]{krall71}
Krall N, Liewer P (1971) Low-frequency instabilities in magnetic pulses. Phys
  Rev A 4(5):2094. \doi{10.1103/PhysRevA.4.2094}

\bibitem[{{Kulsrud} et~al(2005){Kulsrud}, {Ji}, {Fox}, and
  {Yamada}}]{kulsrud05}
{Kulsrud} R, {Ji} H, {Fox} W, et~al (2005) {An electromagnetic drift
  instability in the magnetic reconnection experiment and its importance for
  magnetic reconnection}. Phys Plasmas 12:082,301. \doi{10.1063/1.1949225}

\bibitem[{Le et~al(2015)Le, Egedal, Daughton, Roytershteyn, Karimabadi, and
  Forest}]{le15}
Le A, Egedal J, Daughton W, et~al (2015) {Transition in electron physics of
  magnetic reconnection in weakly collisional plasma}. Journal of Plasma
  Physics 81(01):305810,108. \doi{10.1017/S0022377814000907}

\bibitem[{{Le} et~al(2017){Le}, {Daughton}, {Chen}, and {Egedal}}]{le17}
{Le} A, {Daughton} W, {Chen} LJ, et~al (2017) {Enhanced electron mixing and
  heating in 3-D asymmetric reconnection at the Earth's magnetopause}. Geophys
  Res Lett 44:2096--2104. \doi{10.1002/2017GL072522}

\bibitem[{Le et~al({2016})Le, Egedal, and Daughton}]{le16}
Le A, Egedal J, Daughton W ({2016}) {Two-stage bulk electron heating in the
  diffusion region of anti-parallel symmetric reconnection}. Phys Plasmas
  {23}({10}). \doi{10.1063/1.4964768}

\bibitem[{Lefebvre et~al(2010)Lefebvre, Chen, Gekelman, Kintner, Pickett,
  Pribyl, Vincena, Chiang, and Judy}]{lefebvre10}
Lefebvre B, Chen LJ, Gekelman W, et~al (2010) Laboratory measurements of
  electrostatic solitary structures generated by beam injection. Physical
  Review Letters 105:115,001. \doi{10.1103/PhysRevLett.105.115001}

\bibitem[{Li et~al(2021)Li, Guo, and Liu}]{li21}
Li X, Guo F, Liu YH (2021) The acceleration of charged particles and formation
  of power-law energy spectra in nonrelativistic magnetic reconnection. Phys
  Plasmas 28:052,905. \doi{10.1063/5.0047644}

\bibitem[{Liu et~al(2022)Liu, Cassak, Li, Hesse, Lin, and Genestreti}]{liu22}
Liu YH, Cassak P, Li X, et~al (2022) First-principles theory of the rate of
  magnetic reconnection in magnetospheric and solar plasmas. Communications
  Physics 5. \doi{10.1038/s42005-022-00854-x}

\bibitem[{{Lyons} and {Pridmore-Brown}(1990)}]{lyons90}
{Lyons} LR, {Pridmore-Brown} DC (1990) {Force balance near an X line in a
  collisionless plasma}. J Geophys Res 95:20,903. \doi{10.1029/JA095iA12p20903}

\bibitem[{{Majeski} and {Ji}(2023)}]{majeski23}
{Majeski} S, {Ji} H (2023) {Super-Fermi acceleration in multiscale MHD
  reconnection}. Physics of Plasmas 30(4):042106. \doi{10.1063/5.0139276}

\bibitem[{{Majeski} et~al(2021){Majeski}, {Ji}, {Jara-Almonte}, and
  {Yoo}}]{majeski21}
{Majeski} S, {Ji} H, {Jara-Almonte} J, et~al (2021) {Guide field effects on the
  distribution of plasmoids in multiple scale reconnection}. Physics of Plasmas
  28(9):092106. \doi{10.1063/5.0059017}

\bibitem[{{Matsumoto} et~al(2003){Matsumoto}, {Deng}, {Kojima}, and
  {Anderson}}]{matsumoto03}
{Matsumoto} H, {Deng} XH, {Kojima} H, et~al (2003) {Observation of
  Electrostatic Solitary Waves associated with reconnection on the dayside
  magnetopause boundary}. Geophys Res Lett 30(6):1326.
  \doi{10.1029/2002GL016319}

\bibitem[{Mozer and Pritchett(2011)}]{mozer11}
Mozer FS, Pritchett PL (2011) Electron physics of asymmetric magnetic field
  reconnection. Space Sci Rev 158(1):119--143. \doi{10.1007/s11214-010-9681-8}

\bibitem[{Mozer et~al(2002)Mozer, Bale, and Phan}]{mozer02}
Mozer FS, Bale S, Phan TD (2002) Evidence of diffusion regions at a subsolar
  magnetopause crossing. Phys Rev Lett 89:015,002.
  \doi{10.1103/PhysRevLett.89.015002}

\bibitem[{Mozer et~al(2022)Mozer, Bale, Cattell, Halekas, Vasko, Verniero, and
  Kellogg}]{mozer22}
Mozer FS, Bale SD, Cattell CA, et~al (2022) Core electron heating by triggered
  ion acoustic waves in the solar wind. The Astrophysical Journal Letters
  927:L15. \doi{10.3847/2041-8213/ac5520}

\bibitem[{Ng et~al(2020)Ng, Chen, Le, Stanier, Wang, and Bessho}]{ng20}
Ng J, Chen LJ, Le A, et~al (2020) Lower-hybrid-drift vortices in the
  electron-scale magnetic reconnection layer. Geophysical Research Letters 47.
  \doi{10.1029/2020GL090726}

\bibitem[{Ng et~al(2023)Ng, Yoo, Chen, Bessho, and Ji}]{ng23}
Ng J, Yoo J, Chen LJ, et~al (2023) 3d simulation of lower-hybrid drift waves in
  strong guide field asymmetric reconnection in laboratory experiments. Physics
  of Plasmas 30. \doi{10.1063/5.0138278}

\bibitem[{Norgren et~al(2012)Norgren, Vaivads, Khotyaintsev, and
  Andr√©}]{norgren12}
Norgren C, Vaivads A, Khotyaintsev YV, et~al (2012) Lower hybrid drift waves:
  Space observations. Physical Review Letters 109:055,001.
  \doi{10.1103/PhysRevLett.109.055001}

\bibitem[{{\O}ieroset et~al(2016){\O}ieroset, Phan, Haggerty, Shay, Eastwood,
  Gershman, Drake, Fujimoto, Ergun, Mozer, Oka, Torbert, Burch, Wang, Chen,
  Swisdak, Pollock, Dorelli, Fuselier, Lavraud, Giles, Moore, Saito, Avanov,
  Paterson, Strangeway, Russell, Khotyaintsev, Lindqvist, and
  Malakit}]{oieroset16}
{\O}ieroset M, Phan TD, Haggerty C, et~al (2016) {MMS observations of large
  guide field symmetric reconnection between colliding reconnection jets at the
  center of a magnetic flux rope at the magnetopause}. Geophys Res Lett
  43(1):5536--5544. \doi{10.1002/2016GL069166}

\bibitem[{Oka et~al(2023)Oka, Birn, Egedal, Guo, Ergun, Turner, Khotyaintsev,
  Hwang, Cohen, and Drake}]{oka23}
Oka M, Birn J, Egedal J, et~al (2023) {Particle acceleration by magnetic
  reconnection in geospace}. submitted

\bibitem[{{Olson} et~al(2016){Olson}, {Egedal}, {Greess}, {Myers}, {Clark},
  {Endrizzi}, {Flanagan}, {Milhone}, {Peterson}, {Wallace}, {Weisberg}, and
  {Forest}}]{olson16}
{Olson} J, {Egedal} J, {Greess} S, et~al (2016) {Experimental Demonstration of
  the Collisionless Plasmoid Instability below the Ion Kinetic Scale during
  Magnetic Reconnection}. Physical Review Letters 116(25):255001.
  \doi{10.1103/PhysRevLett.116.255001}

\bibitem[{Olson et~al(2021)Olson, Egedal, Clark, Endrizzi, Greess,
  Millet-Ayala, Myers, Peterson, Wallace, and Forest}]{olson21}
Olson J, Egedal J, Clark M, et~al (2021) Regulation of the normalized rate of
  driven magnetic reconnection through shocked flux pileup. JOURNAL OF PLASMA
  PHYSICS 87(3). \doi{10.1017/S0022377821000659}

\bibitem[{Ono et~al(1993)Ono, Morita, Katsurai, and Yamada}]{ono93}
Ono Y, Morita A, Katsurai M, et~al (1993) Experimental investigation of
  three-dimensional magnetic reconnection by use of two colliding spheromaks.
  Phys Fluids B 5:3691. \doi{10.1063/1.860840}

\bibitem[{Ono et~al(2011)Ono, Tanabe, Hayashi, Ii, Narushima, Yamada, Inomoto,
  and Cheng}]{ono11}
Ono Y, Tanabe H, Hayashi Y, et~al (2011) Ion and electron heating
  characteristics of magnetic reconnection in a two flux loop merging
  experiment. Phys Rev Lett 107(18):185,001.
  \doi{10.1103/PhysRevLett.107.185001}

\bibitem[{Papadopoulos(1977)}]{papadopoulos77}
Papadopoulos K (1977) A review of anomalous resistivity for the ionosphere.
  Reviews of Geophysics and Space Physics 15:113. \doi{10.1029/RG015i001p00113}

\bibitem[{Parker(1957)}]{parker57}
Parker E (1957) Sweet's mechanism for merging magnetic fields in conducting
  fluids. J Geophys Res 62:509. \doi{10.1029/JZ062i004p00509}

\bibitem[{Petschek(1964)}]{petschek64}
Petschek H (1964) Magnetic field annhilation. NASA Spec Pub 50:425

\bibitem[{{Pritchett}(2001)}]{pritchett01}
{Pritchett} PL (2001) {Geospace Environment Modeling magnetic reconnection
  challenge: Simulations with a full particle electromagnetic code}. J Geophys
  Res 106:3783. \doi{10.1029/1999JA001006}

\bibitem[{Pucci and Velli(2014)}]{pucci14}
Pucci F, Velli M (2014) {Reconnection of Quasi-singular Current Sheets: The
  ``Ideal'' Tearing Mode}. Astrophys J 780(2):L19--4.
  \doi{10.1088/2041-8205/780/2/L19}

\bibitem[{{Pucci} et~al(2018){Pucci}, {Usami}, {Ji}, {Guo}, {Horiuchi},
  {Okamura}, {Fox}, {Jara-Almonte}, {Yamada}, and {Yoo}}]{pucci18}
{Pucci} F, {Usami} S, {Ji} H, et~al (2018) {Energy transfer and electron
  energization in collisionless magnetic reconnection for different guide-field
  intensities}. Physics of Plasmas 25(12):122111. \doi{10.1063/1.5050992}

\bibitem[{Ren(2007)}]{renthesis07}
Ren Y (2007) Studies of non-mhd effects during magnetic reconnection in a
  laboratory plasma. PhD thesis, Princeton University

\bibitem[{{Ren} et~al(2005){Ren}, {Yamada}, {Gerhardt}, {Ji}, {Kulsrud}, and
  {Kuritsyn}}]{ren05}
{Ren} Y, {Yamada} M, {Gerhardt} S, et~al (2005) {Experimental Verification of
  the Hall Effect during Magnetic Reconnection in a Laboratory Plasma}. Phys
  Rev Lett 95(5):055,003. \doi{10.1103/PhysRevLett.95.055003}

\bibitem[{{Ren} et~al(2008){Ren}, {Yamada}, Ji, {Gerhardt}, and
  {Kulsrud}}]{ren08}
{Ren} Y, {Yamada} M, Ji H, et~al (2008) {Identification of the Electron
  Diffusion Region during Magnetic Reconnection in a Laboratory Plasma}. Phys
  Rev Lett 101:085,003. \doi{10.1103/PhysRevLett.101.085003}

\bibitem[{{Retin{\`o}} et~al(2021){Retin{\`o}}, {Khotyaintsev}, {Le Contel},
  {Marcucci}, {Plaschke}, {Vaivads}, {Angelopoulos}, {Blasi}, {Burch}, {De
  Keyser}, {Dunlop}, {Dai}, {Eastwood}, {Fu}, {Haaland}, {Hoshino},
  {Johlander}, {Kepko}, {Kucharek}, {Lapenta}, {Lavraud}, {Malandraki},
  {Matthaeus}, {McWilliams}, {Petrukovich}, {Pin{\c{c}}on}, {Saito},
  {Sorriso-Valvo}, {Vainio}, and {Wimmer-Schweingruber}}]{retino21}
{Retin{\`o}} A, {Khotyaintsev} Y, {Le Contel} O, et~al (2021) {Particle
  energization in space plasmas: towards a multi-point, multi-scale plasma
  observatory}. Experimental Astronomy \doi{10.1007/s10686-021-09797-7}

\bibitem[{Roytershteyn et~al(2010)Roytershteyn, Daughton, Dorfman, Ren, Ji,
  Yamada, Karimabadi, Yin, Albright, and Bowers}]{roytershteyn10}
Roytershteyn V, Daughton W, Dorfman S, et~al (2010) Driven reconnection near
  the dreicer limit. Phys Plasmas 17:055,706. \doi{10.1063/1.3399787}

\bibitem[{Roytershteyn et~al(2012)Roytershteyn, Daughton, Karimabadi, and
  Mozer}]{roytershteyn12}
Roytershteyn V, Daughton W, Karimabadi H, et~al (2012) {Influence of the
  Lower-Hybrid Drift Instability on Magnetic Reconnection in Asymmetric
  Configurations}. Phys Rev Lett 108:185,001.
  \doi{10.1103/PhysRevLett.108.185001}

\bibitem[{Roytershteyn et~al(2013)Roytershteyn, Dorfman, Daughton, Ji, Yamada,
  and Karimabadi}]{roytershteyn13}
Roytershteyn V, Dorfman S, Daughton W, et~al (2013) Electromagnetic instability
  of thin reconnection layers: comparison of 3d simulations with mrx
  observations. Phys Plasmas 20:061,212. \doi{10.1063/1.4811371}

\bibitem[{{Russell} and {Elphic}(1979)}]{russell79}
{Russell} CT, {Elphic} RC (1979) {ISEE observations of flux transfer events at
  the dayside magnetopause}. Geophys Res Lett 6:33--36.
  \doi{10.1029/GL006i001p00033}

\bibitem[{Sato and Hayashi(1979)}]{sato79}
Sato T, Hayashi T (1979) Externally driven magnetic reconnection and a powerful
  magnetic energy converter. Phys Fluids 22:1189. \doi{10.1063/1.862721}

\bibitem[{Shay et~al(1998)Shay, Drake, Denton, and Biskamp}]{shay98a}
Shay M, Drake J, Denton R, et~al (1998) Structure of the dissipation region
  during collisionless magnetic reconnection. J Geophys Res 103:9165.
  \doi{10.1029/97JA03528}

\bibitem[{Shi et~al(2022)Shi, Srivastav, Barbhuiya, Cassak, Scime, and
  Swisdak}]{shi22}
Shi P, Srivastav P, Barbhuiya MH, et~al (2022) Laboratory observations of
  electron heating and non-maxwellian distributions at the kinetic scale during
  electron-only magnetic reconnection. Phys Rev Lett 128.
  \doi{10.1103/PhysRevLett.128.025002}

\bibitem[{Sonnerup(1979)}]{sonnerup79}
Sonnerup BU{\"{O}} (1979) Magnetic field reconnection. In: Lanzerotti L, Kennel
  C, Parker E (eds) Solar System Plasma Physics. Cambridge University Press,
  North-Holland New York, p~45

\bibitem[{Stark et~al(2005)Stark, Fox, Egedal, Grulke, and Klinger}]{stark05}
Stark A, Fox W, Egedal J, et~al (2005) Laser-induced fluorescence measurement
  of the ion-energy-distribution function in a collisionless reconnection
  experiment. Phys Rev Lett 95:235,005. \doi{10.1103/PhysRevLett.95.235005}

\bibitem[{Stechow et~al(2018)Stechow, Fox, Jara-Almonte, Yoo, Ji, and
  Yamada}]{stechow18}
Stechow A, Fox W, Jara-Almonte J, et~al (2018) {Electromagnetic fluctuations
  during guide field reconnection in a laboratory plasma}. Phys Plasmas
  25:052,120. \doi{10.1063/1.5025827}

\bibitem[{Steinvall et~al(2021)Steinvall, Khotyaintsev, Graham, Vaivads,
  Andr{\'e}, and Russell}]{steinvall21}
Steinvall K, Khotyaintsev YV, Graham DB, et~al (2021) Large amplitude
  electrostatic proton plasma frequency waves in the magnetospheric separatrix
  and outflow regions during magnetic reconnection. Geophysical Research
  Letters 48(5):e90,286. \doi{10.1029/2020GL090286}

\bibitem[{Stenzel and Gekelman(1981)}]{stenzelgekelman1}
Stenzel R, Gekelman W (1981) Magnetic field line reconnection experiments 1.
  field topologies. J Geophys Res 86:649. \doi{10.1103/PhysRevLett.42.1055}

\bibitem[{{Stenzel} et~al(1986){Stenzel}, {Gekelman}, and
  {Urrutia}}]{stenzel86}
{Stenzel} RL, {Gekelman} W, {Urrutia} JM (1986) {Lessons from laboratory
  experiments on reconnection}. Advances in Space Research 6(1):135--147.
  \doi{10.1016/0273-1177(86)90025-6}

\bibitem[{Stix(1992)}]{stix92}
Stix T (1992) Waves in {P}lasmas. American Institute of Physics, New York

\bibitem[{Sweet(1958)}]{sweet58}
Sweet P (1958) The neutral point theory of solar flares. In: Lehnert B (ed)
  Electromagnetic Phenomena in Cosmical Physics. Cambridge Univ. Press, New
  York, p 123

\bibitem[{Tanabe et~al(2015)Tanabe, Yamada, Watanabe, Gi, Kadowaki, Inomoto,
  Imazawa, Gryaznevich, Michael, Crowley, Conway, Scannell, Harrison,
  Fitzgerald, Meakins, Hawkes, McClements, O'Gorman, Cheng, and Ono}]{tanabe15}
Tanabe H, Yamada T, Watanabe T, et~al (2015) Electron and ion heating
  characteristics during magnetic reconnection in the mast spherical tokamak.
  Phys Rev Lett 115:215,004. \doi{10.1103/PhysRevLett.115.215004}

\bibitem[{Terasawa(1983)}]{terasawa83}
Terasawa T (1983) Hall current effect on tearing mode instability. Geophys Res
  Lett 10:475. \doi{10.1029/GL010i006p00475}

\bibitem[{{Torbert} et~al(2016){Torbert}, {Burch}, {Giles}, {Gershman},
  {Pollock}, {Dorelli}, {Avanov}, {Argall}, {Shuster}, {Strangeway}, {Russell},
  {Ergun}, {Wilder}, {Goodrich}, {Faith}, {Farrugia}, {Lindqvist}, {Phan},
  {Khotyaintsev}, {Moore}, {Marklund}, {Daughton}, {Magnes}, {Kletzing}, and
  {Bounds}}]{torbert16}
{Torbert} RB, {Burch} JL, {Giles} BL, et~al (2016) {Estimates of terms in Ohm's
  law during an encounter with an electron diffusion region}. Geophys Res Lett
  43:5918--5925. \doi{10.1002/2016GL069553}

\bibitem[{{Torbert} et~al(2018){Torbert}, {Burch}, {Phan}, {Hesse}, {Argall},
  {Shuster}, {Ergun}, {Alm}, {Nakamura}, {Genestreti}, {Gershman}, {Paterson},
  {Turner}, {Cohen}, {Giles}, {Pollock}, {Wang}, {Chen}, {Stawarz}, {Eastwood},
  {Hwang}, {Farrugia}, {Dors}, {Vaith}, {Mouikis}, {Ardakani}, {Mauk},
  {Fuselier}, {Russell}, {Strangeway}, {Moore}, {Drake}, {Shay},
  {Khotyaintsev}, {Lindqvist}, {Baumjohann}, {Wilder}, {Ahmadi}, {Dorelli},
  {Avanov}, {Oka}, {Baker}, {Fennell}, {Blake}, {Jaynes}, {Le Contel},
  {Petrinec}, {Lavraud}, and {Saito}}]{torbert18}
{Torbert} RB, {Burch} JL, {Phan} TD, et~al (2018) {Electron-scale dynamics of
  the diffusion region during symmetric magnetic reconnection in space}.
  Science 362(6421):1391--1395. \doi{10.1126/science.aat2998}

\bibitem[{{Trivelpiece} and {Gould}(1959)}]{trivelpiece59}
{Trivelpiece} AW, {Gould} RW (1959) {Behavior of Polytetrafluoroethylene
  (Teflon) under High Pressures}. Journal of Applied Physics 30(11):1784--1793.
  \doi{10.1063/1.1735056}

\bibitem[{Uchino et~al(2017)Uchino, Kurita, Harada, Machida, and
  Angelopoulos}]{uchino17}
Uchino H, Kurita S, Harada Y, et~al (2017) Waves in the innermost open boundary
  layer formed by dayside magnetopause reconnection. Journal of Geophysical
  Research: Space Physics 122:3291--3307. \doi{10.1002/2016JA023300}

\bibitem[{Ugai and Tsuda(1977)}]{ugai77}
Ugai M, Tsuda T (1977) Magnetic field-line reconnection by localized
  enhancement of reconnection. 1. evolution in a compressible mhd fluid. J
  Plasma Phys 17:337. \doi{10.1017/S0022377800020663}

\bibitem[{Uzdensky et~al(2010)Uzdensky, Loureiro, and
  Schekochihin}]{uzdensky10}
Uzdensky DA, Loureiro NF, Schekochihin A (2010) Fast magnetic reconnection in
  the plasmoid-dominated regime. Phys Rev Lett 105:235,002.
  \doi{10.1103/PhysRevLett.105.235002}

\bibitem[{Vasyliunas(1975)}]{vasyliunas75}
Vasyliunas V (1975) Theoretical models of field line merging, i. Rev Geophys
  Space Phys 13:303. \doi{10.1029/RG013i001p00303}

\bibitem[{Wilder et~al(2018)Wilder, Ergun, Burch, Ahmadi, Eriksson, Phan,
  Goodrich, Shuster, Rager, Torbert, Giles, Strangeway, Plaschke, Magnes,
  Lindqvist, and Khotyaintsev}]{wilder18}
Wilder FD, Ergun RE, Burch JL, et~al (2018) The role of the parallel electric
  field in electron-scale dissipation at reconnecting currents in the
  magnetosheath. Journal of Geophysical Research: Space Physics 123:6533--6547.
  \doi{10.1029/2018JA025529}

\bibitem[{Wygant et~al(2005)Wygant, Cattell, Lysak, Song, Dombeck, McFadden,
  Mozer, Carlson, Parks, Lucek, Balogh, Andre, Reme, Hesse, and
  Mouikis}]{wygant05}
Wygant J, Cattell C, Lysak R, et~al (2005) Cluster observations of an intense
  normal component of the electric field at a thin reconnecting current sheet
  in the tail and its role in the shock-like acceleration of the ion fluid into
  the separatrix region. J Geophys Res 110:A09,206. \doi{10.1029/2004JA010708}

\bibitem[{Yamada(2022)}]{yamada22}
Yamada M (2022) Magnetic Reconnection: A Modern Synthesis of Theory,
  Experiment, and Observations. Princeton University Press, Princeton

\bibitem[{Yamada et~al(1990)Yamada, Ono, Hayakawa, Katsurai, and
  Perkins}]{yamada90}
Yamada M, Ono Y, Hayakawa A, et~al (1990) Magnetic reconnection of plasma
  toroids with co- and counter-helicity. Phys Rev Lett 65:721.
  \doi{10.1103/PhysRevLett.65.721}

\bibitem[{Yamada et~al(1997)Yamada, Ji, Hsu, Carter, Kulsrud, Bretz, Jobes,
  Ono, and Perkins}]{yamada97b}
Yamada M, Ji H, Hsu S, et~al (1997) Study of driven magnetic reconnection in a
  laboratory plasma. Phys Plasmas 4:1936. \doi{10.1063/1.872336}

\bibitem[{{Yamada} et~al(2006){Yamada}, {Ren}, {Ji}, {Breslau}, {Gerhardt},
  {Kulsrud}, and {Kuritsyn}}]{yamada06}
{Yamada} M, {Ren} Y, {Ji} H, et~al (2006) {Experimental study of two-fluid
  effects on magnetic reconnection in a laboratory plasma with variable
  collisionality}. Phys Plasmas 13:052,119. \doi{10.1063/1.2203950}

\bibitem[{Yamada et~al(2010)Yamada, Kulsrud, and Ji}]{yamada10}
Yamada M, Kulsrud R, Ji H (2010) Magnetic reconnection. Rev Mod Phys 82:603.
  \doi{10.1103/RevModPhys.82.603}

\bibitem[{Yamada et~al(2014)Yamada, Yoo, Jara-Almonte, Ji, Kulsrud, and
  Myers}]{yamada14}
Yamada M, Yoo J, Jara-Almonte J, et~al (2014) Conversion of magnetic energy in
  the magnetic reconnection layer of a laboratory plasma. Nat Commun 5:4774.
  \doi{10.1038/ncomms5774}

\bibitem[{{Yamada} et~al(2015){Yamada}, {Yoo}, {Jara-Almonte}, {Daughton},
  {Ji}, {Kulsrud}, and {Myers}}]{yamada15}
{Yamada} M, {Yoo} J, {Jara-Almonte} J, et~al (2015) {Study of energy conversion
  and partitioning in the magnetic reconnection layer of a laboratory
  plasmas)}. Phys Plasmas 22(5):056501. \doi{10.1063/1.4920960}

\bibitem[{Yamada et~al(2016)Yamada, Yoo, and Myers}]{yamada16}
Yamada M, Yoo J, Myers CE (2016) Understanding the dynamics and energetics of
  magnetic reconnection in a laboratory plasma: Review of recent progress on
  selected fronts. Phys Plasmas 23(5):055,402. \doi{10.1063/1.4948721}

\bibitem[{Yamada et~al(2018)Yamada, Chen, Yoo, Wang, Fox, Jara-Almote, Ji,
  Daughton, Le, Burch, Giles, Hesse, Moore, and Torbert}]{yamada18}
Yamada M, Chen LJ, Yoo J, et~al (2018) The two-fluid dynamics and energetics of
  the asymmetric magnetic reconnection in laboratory and space plasmas. Nat
  Commun 8:5223. \doi{10.1038/s41467-018-07680-2}

\bibitem[{Yoo et~al(2013)Yoo, Yamada, Ji, and Myers}]{yoo13}
Yoo J, Yamada M, Ji H, et~al (2013) Observation of ion acceleration and heating
  during collisionless magnetic reconnection in a laboratory plasma. Phys Rev
  Lett 110:215,007. \doi{10.1103/PhysRevLett.110.215007}

\bibitem[{Yoo et~al(2014{\natexlab{a}})Yoo, Yamada, Ji, Jara-Almonte, and
  Myers}]{yoo14b}
Yoo J, Yamada M, Ji H, et~al (2014{\natexlab{a}}) Bulk ion acceleration and
  particle heating during magnetic reconnection in a laboratory plasma. Phys
  Plasmas 21(5):055,706. \doi{10.1063/1.4874331}

\bibitem[{Yoo et~al(2014{\natexlab{b}})Yoo, Yamada, Ji, Jara-Almonte, Myers,
  and Chen}]{yoo14}
Yoo J, Yamada M, Ji H, et~al (2014{\natexlab{b}}) {Laboratory Study of Magnetic
  Reconnection with a Density Asymmetry across the Current Sheet}. Phys Rev
  Lett 113(9):095,002. \doi{10.1103/PhysRevLett.113.095002}

\bibitem[{Yoo et~al(2017)Yoo, Na, Jara-Almonte, Yamada, Ji, Roytershteyn, Fox,
  and Chen}]{yoo17}
Yoo J, Na B, Jara-Almonte J, et~al (2017) {Electron heating and energy
  inventory during asymmetric reconnection in a laboratory plasma}. J Geophys
  Res 122:9264--9281. \doi{10.1002/2017JA024152}

\bibitem[{{Yoo} et~al(2018){Yoo}, {Jara-Almonte}, {Yerger}, {Wang}, {Qian},
  {Le}, {Ji}, {Yamada}, {Fox}, {Kim}, {Chen}, and {Gershman}}]{yoo18}
{Yoo} J, {Jara-Almonte} J, {Yerger} E, et~al (2018) {Whistler Wave Generation
  by Anisotropic Tail Electrons During Asymmetric Magnetic Reconnection in
  Space and Laboratory}. Geophys Res Lett 45(16):8054--8061.
  \doi{10.1029/2018GL079278}

\bibitem[{{Yoo} et~al(2019){Yoo}, {Wang}, {Yerger}, {Jara-Almonte}, {Ji},
  {Yamada}, {Chen}, {Fox}, {Goodman}, and {Alt}}]{yoo19}
{Yoo} J, {Wang} S, {Yerger} E, et~al (2019) {Whistler wave generation by
  electron temperature anisotropy during magnetic reconnection at the
  magnetopause}. Physics of Plasmas 26(5):052902. \doi{10.1063/1.5094636}

\bibitem[{{Yoo} et~al(2020){Yoo}, {Ji}, {Ambat}, {Wang}, {Ji}, {Lo}, {Li},
  {Ren}, {Jara-Almonte}, {Chen}, {Fox}, {Yamada}, {Alt}, and {Goodman}}]{yoo20}
{Yoo} J, {Ji} JY, {Ambat} MV, et~al (2020) {Lower Hybrid Drift Waves During
  Guide Field Reconnection}. Geophys Res Lett 47(21):e87192.
  \doi{10.1029/2020GL087192}

\bibitem[{Yoo et~al(2023)Yoo, Ng, Ji, Bose, Goodman, Alt, Chen, Shi, and
  Yamada}]{yoo23}
Yoo J, Ng J, Ji H, et~al (2023) {Anomalous resistivity and electron heating by
  lower hybrid drift waves during magnetic reconnection with a guide field}.
  submitted

\bibitem[{{Zenitani} and {Hoshino}(2001)}]{zenitani01}
{Zenitani} S, {Hoshino} M (2001) {The Generation of Nonthermal Particles in the
  Relativistic Magnetic Reconnection of Pair Plasmas}. Astrophys J Lett
  562:L63--L66. \doi{10.1086/337972}

\bibitem[{Zenitani et~al(2011)Zenitani, Hesse, Klimas, and
  Kuznetsova}]{zenitani11}
Zenitani S, Hesse M, Klimas A, et~al (2011) New measure of the dissipation
  region in collisionless magnetic reconnection. Phys Rev Lett 106.
  \doi{10.1103/PhysRevLett.106.195003}

\bibitem[{{Zhang} et~al(2023){Zhang}, {Chien}, {Gao}, {Ji}, {Blackman},
  {Follett}, {Froula}, {Katz}, {Li}, {Birkel}, {Petrasso}, {Moody}, and
  {Chen}}]{zhang23}
{Zhang} S, {Chien} A, {Gao} L, et~al (2023) {Ion and Electron Acoustic Bursts
  during Anti-Parallel Reconnection Driven by Lasers}. Nature Physics
  \doi{10.1038/s41567-023-01972-1}

\bibitem[{{Zhou} et~al(2009){Zhou}, {Deng}, {Li}, {Pang}, {Vaivads},
  {R{\`e}me}, {Lucek}, {Fu}, {Lin}, {Yuan}, and {Wang}}]{zhou09}
{Zhou} M, {Deng} XH, {Li} SY, et~al (2009) {Observation of waves near lower
  hybrid frequency in the reconnection region with thin current sheet}. J
  Geophys Res 114:2216. \doi{10.1029/2008JA013427}

\bibitem[{Zweibel and Yamada(2009)}]{zweibel09}
Zweibel E, Yamada M (2009) Magnetic reconnection in astrophysical and
  laboratory plasmas. Annu Rev Astron Astrophys 47(1):291.
  \doi{10.1146/annurev-astro-082708-101726}

\end{thebibliography}


\end{document}